\documentclass[a4paper,11pt]{article}

\usepackage{jcappub} 

\usepackage[T1]{fontenc} 

\usepackage{amssymb}

\newcommand{\C}{\mathbf C}
\newcommand{\Y}{\mathbf Y}
\newcommand{\0}{\mathbf 0}
\newcommand{\F}{\mathbf F}
\newcommand{\M}{\mathbf M}
\newcommand{\D}{\mathbf D}
\newcommand{\N}{\mathbf N}

\newcommand{\lm}{\ell_{\max}}

\newcommand{\x}{\mathbf{x}}

\newcommand{\mP}{\mathbf{P}}
\newcommand{\mS}{\mathbf{S}}
\newcommand{\vvec}[1]{\mathbf{#1}}

\newcommand{\SU}{S_{I}}
\newcommand{\SD}{S_{II}}
\newcommand{\ST}{S_{III}}
\newcommand{\pr}{\mathrm{res}}

\DeclareMathOperator{\rank}{rank}

\def\refjnl#1{{\rmfamily #1}}
\newcommand\apj{\refjnl{ApJ}}  		
\newcommand\apjl{\refjnl{ApJL}}     
\newcommand\aap{\refjnl{A\&A}}	    
\newcommand\mnras{\refjnl{MNRAS}}   
\newcommand\prd{\refjnl{PhRvD}} 	
\newcommand\jcap{\refjnl{JCAP}}  	

\title{On the regularity of the covariance matrix of a discretized scalar field on the sphere}

\author[a,b]{J.~D. Bilbao-Ahedo}
\author[b]{R.~B. Barreiro}
\author[b]{D. Herranz}
\author[b]{P. Vielva}
\author[b]{E. Mart\'{\i}nez-Gonz\'alez}

\affiliation[a]{Departamento de F\'{\i}sica Moderna, Universidad de Cantabria,\\ Av.\ los Castros s/n, 39005 Santander (Spain)}
\affiliation[b]{Instituto de F\'{\i}sica de Cantabria (CSIC-UC),\\ Av.\ los Castros s/n, 39005 Santander (Spain)}

\emailAdd{bilbao@ifca.unican.es}
\emailAdd{barreiro@ifca.unican.es}
\emailAdd{herranz@ifca.unican.es}
\emailAdd{vielva@ifca.unican.es}
\emailAdd{martinez@ifca.unican.es}

\abstract{We present a comprehensive study of the regularity of the covariance matrix of a discretized field on the sphere. In a particular situation, the rank of the matrix depends on the number of pixels, the number of spherical harmonics, the symmetries of the pixelization scheme and the presence of a mask. Taking into account the above mentioned components, we provide analytical expressions that constrain the rank of the matrix. They are obtained by expanding the determinant of the covariance matrix as a sum of determinants of matrices made up of spherical harmonics. We investigate these constraints for five different pixelizations that have been used in the context of Cosmic Microwave Background (CMB) data analysis: \texttt{Cube}, \texttt{Icosahedron}, \texttt{Igloo}, \texttt{GLESP} and \texttt{HEALPix}, finding that, at least in the considered cases, the \texttt{HEALPix} pixelization tends to provide a covariance matrix with a rank closer to the maximum expected theoretical value than the other pixelizations. The effect of the propagation of numerical errors in the regularity of the covariance matrix is also studied for different computational precisions, as well as the effect of adding a certain level of noise in order to regularize the matrix. In addition, we investigate the application of the previous results to a particular example that requires the inversion of the covariance matrix: the estimation of the CMB temperature power spectrum through the Quadratic Maximum Likelihood algorithm. Finally, some general considerations in order to achieve a regular covariance matrix are also presented.}

\keywords{CMBR theory, CMBR experiments.}

\begin{document}
\maketitle
\flushbottom

\section{Introduction}

With the advent of the precision era of Cosmology and `big data' astrophysical experiments, data analysis techniques have risen to a prominent role in modern Astronomy. In particular, the treatment of very large matrices is a challenge from the point of view of algebraic and numerical methods, data storage and software implementation. For example, many astrophysical problems require the manipulation of large covariance matrices and their inverses. But it is often the case that such matrices are ill-conditioned and their inverse matrices cannot be calculated; when this happens, the problem must be attacked either by means of clever algorithms that calculate the pseudo-inverse of the matrix \citep[see, e.g.,][]{pseudo} or by \emph{ad hoc} regularizers such as the addition of a small amount of uncorrelated noise to the diagonal of the covariance matrix. 

The literature is rich in situations where the regularity of the covariance matrix plays a fundamental role. For example, the inverse of the covariance matrix is necessary for the study of the statistics of the Cosmic Microwave Background (CMB), the Quadratic Maximum Likelihood (QML) power spectrum estimator \cite{QML}, the maximum likelihood cosmological parameter estimation~\cite{Planck2013_XV,Planck2015_XI}, to study the topology of the universe~\cite{2013JCAP...08..009A} and for many CMB foreground removal/component separation methods \citep[see, e.g.,][]{ILC,eriksen04,eriksen07,2014A&A...571A..12P}. Beyond the mere characterization of the second-order statistics of the CMB temperature or polarization fluctuations, covariance matrices are also fundamental for the study of non-Gaussianity \cite{vielva09, curto11} and the statistical analysis of CMB anomalies such as the Cold Spot \cite{cruz07}. But the covariance matrix is often ill-conditioned even for low resolution sky maps.

The typical solution consists on the regularization `by hand' of the covariance matrix. For example a small level of white noise is added to the noise covariance matrix of the CMB temperature data involved in the construction of the low-multipole Planck likelihood~\cite{Planck2015_XI}. Another possibility is to deal with ill-conditioned covariance matrices by using a principal component analysis approach to remove the lowest degenerate eigenvalues, as is done for instance to study the multi-normality of the CMB~\cite{2016A&A...594A..16P} or in the estimation of primordial non-Gaussianity using wavelets~\cite{curto11}. The main problem with this kind of approaches is the \emph{ad hoc} nature of the regularization. For example, the amount of uncorrelated artificial noise to be added to the covariance matrix must be carefully chosen: if the level of the noise is too small, it may not suffice to make the matrix regular, but if it is too large, the quality of the data is sorely compromised. One must find the correct amount of noise by trial and error.

A more fundamental question is why and how covariance matrices become ill-conditioned. By definition, all covariance matrices should be positive-semidefinite and symmetric. Therefore, a covariance matrix can be singular if it has, at least, one eigenvalue equal to zero. However, in practice it is often assumed that if a covariance matrix arising from CMB data is singular, it must be so because of numerical issues related to the way the inverse is computed and to limits in computer precision. In this paper we will focus on the interesting case of observations on the sphere, and we will make a comprehensive study on how the regularity and rank of the covariance matrix depend on how it is built, the pixelization scheme, the sky coverage and the regularizing noise. We will show that, even with arbitrary high precision, CMB covariance matrices can be singular due to the way the CMB is sampled. As we will see in the following sections, the way the sphere is pixelised  can introduce symmetries that affect the regularity of covariance matrices. This is a purely algebraic effect that, as far as we know, has not been explored in the literature before.

The structure of the paper is as follows: section~\ref{sec:CondicionNecesaria} enunciates a general constraint in the rank of the covariance matrix. Section~\ref{sec:Simetrias} presents further upper limits on the rank under the presence of some specific symmetries of the pixelization. These constraints are tested for five different pixelization schemes in section~\ref{sec:Pixelizaciones}. Section~\ref{sec:Espectro} studies the effect of introducing a realistic CMB model when computing the covariance matrix as well as the degrading effect of numerical precision. Section~\ref{sec:Ruido} investigates the addition of noise as a regulariser of the covariance matrix. The effect of the presence of a mask in the rank of the covariance matrix is addressed in section~\ref{sec:Mascara}. As an example, section~\ref{sec:QML} applies the results of the previous sections to the Quadratic Maximum Likelihood method~\cite{QML} for the estimation of the angular power spectrum, which requires the calculation of the inverse of the covariance matrix. Concluding remarks are offered in section~\ref{sec:Conclusiones}. Finally, technical points related to the covariance matrix are presented in the appendices.

\section{The rank of the covariance matrix}
\label{sec:CondicionNecesaria}

If a discretized scalar field on the sphere arises from isotropic random fluctuations, the elements of the covariance matrix $\C \equiv \langle \x \x^t \rangle$ can be written in terms of the angular power spectrum $C_{\ell}$ as:
\begin{equation}
\label{eq:eq1}
\C_{ij}= \sum_{\ell= 0}^{\infty} C_{\ell}\sum_{m=-\ell}^{\ell}  Y_{\ell m}(\theta_i, \phi_i) Y^*_{\ell m}(\theta_j, \phi_j),
\end{equation}
where $Y_{\ell m}$ are the spherical harmonics given by:
\begin{equation}
Y_{\ell m}(\theta, \phi)=\sqrt{\frac{2\ell+1}{4\pi}\frac{(l-m)!}{(l+m)!}}P_{\ell
m}(\cos{\theta})e^{i m \phi},
\end{equation}
\noindent and $P_{\ell m}$ are the associated Legendre polynomials. In
real life applications, the sum over multipoles does not extend to infinity, but it has a cutoff at some $\lm$ instead. If the value of $\lm$ is not sufficiently large, the covariance matrix can be singular. This is a problem for all numerical applications that require the inversion of $\C$. On the other hand, a too large value of $\lm$ can lead to extremely heavy computational costs. Besides, there is a limit in the spatial resolution that can be explored, which is determined by the lowest angular distance between points in the pixelization, as established by the Nyquist's theorem. Therefore, one must be very careful with the choice of $\lm$. In this section we will make a thorough study of the rank of $\C$ as a function of $\lm$.

Let $\ell_{\min}$ and $\lm$ be the summation limits in the finite version of eq.~\eqref{eq:eq1}. Let us also consider a pixelization of the sky such that the unit sphere is sampled at a set of $n$ positions $\lbrace \vvec{r}_i \rbrace$ and angular coordinates $\lbrace \left( \theta_i, \phi_i \right) \rbrace$, $i = 1, \ldots, n$. In order to simplify the notation in the following expressions, we introduce the univocal index change $ \left( \ell, m \right) \leftrightarrow \mu $ for the spherical harmonics such that:
\begin{equation}
\mu \left( \ell, m \right) = \sum_{k = \ell_{\min}}^{\ell-1} \left( 2 k +1 \right) + \ell+ m +1 = \ell^2 -\ell_{\min}^2 + \ell+ m +1 .
\end{equation}
\noindent This change assigns a unique $\mu$ to each pair $(\ell,m)$ in
ascending $m$ order, that is, $\mu(\ell_{\min},-\ell_{\min}) = 1$,
$\mu(\ell_{\min},-\ell_{\min}+1) = 2$ and so on. $\mu$ runs from 1 to $N$,
the total number of spherical harmonics, given by:
\begin{equation}
\label{TotalArmonicos}
N = \sum_{\ell=\ell_{\min}}^{\ell_{\max}}{(2 \ell+1)} =\ell_{\max}^2 - \ell_{\min}^2 + 2 \ell_{\max} + 1.
\end{equation}
We will also use the same index $\mu$ for the power spectrum, such that $C_\mu$ corresponds to the power of the multipole $\ell$ obtained as:
\begin{equation}
\ell= \mathrm{floor} \left( \sqrt{\ell_{\min}^2-1+\mu} \right),
\end{equation}
where $\mathrm{floor(x)}$ is the largest integer not greater than $\mathrm{x}$.
Using this notation, we can write:
\begin{equation}
\label{CambioIndices}
Y_{\ell m}(\theta_i, \phi_{i}) = Y_{\mu}(\theta_i, \phi_{i}) = Y_{\mu i},
\end{equation}
and thus:
\begin{equation}
\label{Sumatorio}
\C_{ij} = \sum_{\ell= \ell_{\min}}^{\ell_{\max}} C_{\ell}\sum_{m=-\ell}^{\ell}Y_{\ell m}(\theta_i, \phi_i) Y^*_{\ell m}(\theta_j, \phi_j) = \sum_{\mu=1}^{N} C_\mu Y_{\mu i} Y_{\mu j}^*,
\end{equation}
\noindent where $\mu$ makes reference to the harmonic indexes, and $i$ and $j$ make reference to the pair of points on the sky at which the sum is being calculated. Using this notation, the $j^{\mathrm{th}}$ vector column of $\C$ is:
\begin{equation}
\label{Columna}
\C_{\texttt{Col}_j} = \sum_{\mu_j} C_{\mu{_j}} \left(  \begin{array}{c} 
Y_{\mu_j 1} \\ Y_{\mu_j 2} \\ \vdots \\ Y_{\mu_j n}
\end{array}\right) Y_{\mu_j j}^*,
\end{equation}
\noindent and therefore $\C$ is:
\begin{equation}
\label{MatrizC}
\C = \left(  
\begin{array}{ccc}
\sum_{\mu_1} C_{\mu_1} \left( 
\begin{array}{c} 
Y_{\mu_1 1} \\ \vdots \\ Y_{\mu_1 n} 
\end{array}  \right) Y_{\mu_1 1}^*  & \cdots &
\sum_{\mu_{n}} C_{\mu_{n}} \left(  \begin{array}{c} 
Y_{\mu_n 1} \\ \vdots \\ Y_{\mu_n n}
\end{array} \right) Y_{\mu_n n}^*
\end{array}
\right).
\end{equation}
Using eq.~\eqref{MatrizC} and the properties of matrix determinants, we can write the determinant of $\C$ as:
\begin{equation}
\label{FinalDesarrollo}
|\C| = \sum_{\mu_1 \cdots \mu_n = 1}^{N} \left( \prod_{i=1}^{n} C_{\mu_i}
Y_{\mu_i i}^* \right)
\left| 
\begin{array}{ccc}
Y_{\mu_1 1}   & \cdots & Y_{\mu_n 1} \\
\vdots & & \vdots \\
Y_{\mu_1 n}  & \cdots & Y_{\mu_n n} \\
\end{array}
\right|.
\end{equation}

We are interested in knowing whether or not $|\C|=0$. A sufficient condition for this determinant to be zero is that all the determinants of the previous equation are null.

If we calculate $\C$ on $n$ points on the sphere using $N$ spherical harmonics, the number of elements of the sum given in eq.~\eqref{FinalDesarrollo} is $N^n$. Any of those determinants can be different from zero if its columns are linearly independent. Since the necessary condition for this to be achieved is that each column corresponds to a different spherical harmonic, we have a first constraint on $\lm$: $N$ must be equal or greater than $n$. In other words, the number of spherical harmonics must be at least as large as the number of considered pixels on the sphere. The rank of the covariance matrix $\C$ is thus constrained by:
\begin{equation}
\label{RangoSinSimetria}
\rank (\C) \le \min \left( n, \ell_{\max}^2 - \ell_{\min}^2 + 2 \ell_{\max} + 1 \right).
\end{equation}
Hereinafter we will refer to this constraint as $R_0$. 

According to eq.~\eqref{FinalDesarrollo}, if one determinant in the sum is not null, there will be other $n!-1$ terms that are not zero as well, which correspond to the permutations of its columns. However, we may wonder if the sum of all these elements could be zero. In appendix~\ref{Determinante} we show that the sum of these terms is a real positive number and, therefore, if there exists at least one non-null determinant, then $\left|\C\right| > 0$ is satisfied. Nonetheless, note that having $N \ge n$ does not guarantee that this is fulfilled. In particular, even if all the columns of the determinants correspond to different spherical harmonics, they can still be linearly dependent. A particular case when this can happen is under the presence of certain symmetries in the considered pixelization of the sphere. This is studied in more detail in the next section.

\section{\texorpdfstring{The effect of symmetries on the rank of \boldmath $\C$}{The effect of symmetries on the rank of C}}
\label{sec:Simetrias}

It is possible that the rank of $\C$ is reduced if the positions where the covariance matrix is calculated have certain symmetries. In this section we will study several of these symmetries, which appear in some commonly used pixelizations of the sphere, and their effect on the rank of $\C$.

As detailed in section~\ref{sec:CondicionNecesaria}, the determinant of $\C$ can be expanded as the sum of the determinants of several matrices according to eq.~\eqref{FinalDesarrollo}. It is easy to see that many of these matrices will have a null determinant but, if the sum runs up to a sufficiently large $\ell$, some of them might have non-zero determinants. Rather than inspecting the rank of each individual matrix, it is possible to determine the existence (or non existence) of full-rank matrices in eq.~\eqref{FinalDesarrollo} by inspecting the (non necessarily square) matrix, whose elements are given by:
\begin{equation}
\label{MatrizGenericaB}
\Y = \left(  
\begin{array}{ccc}
Y_{1,1}  & \cdots & Y_{N,1} \\
\vdots & & \vdots \\
Y_{1,n} & \cdots & Y_{N,n} \\
\end{array} \right),
\end{equation}
\noindent
where the harmonics are calculated for all the $n$ pixels and $N$ is
the number of spherical harmonics. Therefore, $\Y$ is an $n\times N$ matrix, with as many rows as
pixels and as many columns as spherical harmonics. For convenience, the value of the element $ij$ of $\Y$ is the value of the spherical harmonic of index $\mu=j$ on the pixel $\vvec{r}_i$, $\Y_{ij}=Y_{ji}=Y_{j}(\vvec{r}_i)$.

It is easy to see that $\rank(\C)=\rank(\Y)$. In particular, if $N \ge
n$, the maximum possible rank of $\Y$ will be n. If this rank is
achieved, this implies that there exists at least $n$ linearly
independent columns in $\Y$ and, thus, there is at least one
determinant in eq.~\eqref{FinalDesarrollo} together with its
permutations which are different from zero.

With the help of the matrix $\Y$ and defining a diagonal signal matrix $\mathbf{S}$, $\mathbf{S}_{\mu \nu}=C_{\mu} \delta_{\mu \nu}$, we can get to the same conclusion by a different route. The expression~\eqref{Sumatorio} can be written as a product of matrices, $\C = \Y \mathbf{S} \Y^{\dag}$. Since the rank of the product of matrices is lower than or equal to the minimum rank of the factors, for $\C$ to be regular the number of columns of $\Y$ (the number of spherical harmonics) has to be equal to or greater than the dimensions of $\C$ (the number of pixels).\footnote{This approach can be useful to analyse the rank of the polarization covariance matrix. In this case, the matrix $\Y$ is composed of spherical harmonics of spin 0 and combinations of harmonics of spin $\pm$2 ~\citep{1998ApJ...503....1Z}. The number of rows of $\Y$ is $3 \times n$ and the number of columns is $3 \times N$. Therefore, $\lm$ has to be so that $N \ge n$.}   

In the next subsections we will study how the rank of $\Y$ is affected
by some specific symmetries.

\subsection{\texorpdfstring{\boldmath $\SU$ symmetry: $\vvec{r} \rightarrow -\vvec{r}$}{SI symmetry}}
\label{sec:s1}
First of all, we will consider a pixelization such that all points on
the sphere have a diametrically opposed point. For full sky coverage,
one would expect that this condition is in general fulfilled since it
is reasonable that two symmetrical hemispheres are pixelised in the
same way. However, this may not be the case when only partial coverage
of the sky is considered.

Let us make the variable change $\cos \theta = z$ in $Y_{\ell m} (\theta,
\phi)$. The spherical
harmonics $Y_{\ell m} (z, \phi)$ have a well-defined parity with respect
to the change $(z,\phi) \rightarrow (-z,\phi + \pi)$ 
(i.e. $\vvec{r} \rightarrow -\vvec{r}$):
\begin{equation}
\label{SimetriaY_Texto}
Y_{\ell m}(-z, \phi + \pi) = {(-1)}^{\ell}Y_{\ell m}(z, \phi).
\end{equation}
Assuming that we have a pixelization and sky coverage such that 
for each pixel with direction $\vvec{r}$ there exists a pixel in the
direction $-\vvec{r}$, eq.~\eqref{SimetriaY_Texto} allows us to transform the matrix 
of eq.~\eqref{MatrizGenericaB} into a block-diagonal matrix with the same rank as the matrix $\Y$:
\begin{equation}
\label{BloquesY1}
\Y \rightarrow
\left(\begin{array}{cc} \Y_e & \0 \\ \0 & \Y_o\end{array}\right),
\end{equation}
\noindent
where $\Y_e$ is a block of  spherical harmonics of even $\ell$
calculated on half of the points of the pixelization, and $\Y_o$
contains the spherical harmonics with odd $\ell$ (see
appendix~\ref{Simetria1} for details).

Therefore, we can write $\rank (\Y) = \rank (\Y_e)+\rank (\Y_o)$. Since
the rank of each block is limited by the minimum value of the number
of rows and the number of columns, the following constraint must be
satisfied:
\begin{equation}\label{FINAL_Texto}
\rank (\Y)  \le  \min \left(N_e, n/2 \right) + \min \left(N_o, n/2 \right),
\end{equation}
\noindent
where
\begin{equation}
\label{eq:eses}
N_e = \sum_{ \substack{\ell=\ell_{\min} \\ (\ell\; even) }}^{\ell_{\max}} (2\ell+1), \qquad
N_o = \sum_{ \substack{\ell=\ell_{\min} \\ (\ell\; odd) }}^{\ell_{\max}} (2\ell+1).
\end{equation}
Hereinafter we will refer to the relationship~\eqref{FINAL_Texto}
as $R_1$. As will be shown below, this additional constraint can
lead in certain cases to the reduction of the maximum rank that the
covariance matrix can achieve. In particular, this could yield to
a singular covariance matrix even in the case $N\ge n$.

\subsection{\texorpdfstring{\boldmath $\SD$ symmetry: $\phi \rightarrow \phi+\pi$}{SII symmetry}}
\label{sec:s2}

Let us assume that we have a pixelization that, in addition to the
previous symmetry $\SU$, satisfies that for each point with coordinates
$(z,\phi)$ there is a point with coordinates
$(z,\phi+\pi)$.

Taking into account the following expression of the spherical harmonics:
\begin{equation}
\label{AESimetriaMVez1}
Y_{\ell m}(z, \phi+\pi)  = {(-1)}^m Y_{\ell m}(z, \phi),
\end{equation}
we can further expand the matrix $\Y$ in blocks (see
appendix~\ref{Simetria2} for details):
\begin{equation}
\label{BloquesY2}
\Y \rightarrow
\left(\begin{array}{cc} 
\begin{array}{cc} \Y_{eo} & \0 \\ \0 &  \Y_{ee} \end{array}  & \ \0 \\ 
\0 & \begin{array}{cc} \Y_{oo} & \0 \\ \0 & \Y_{oe} \end{array}
\end{array}\right),
\end{equation}
\noindent
where, for example, $\Y_{eo}$ denotes the block made up of the spherical harmonics of even $\ell$ and odd $m$.  The size of the
blocks depends on the number of pixels at $z=0$ in the
pixelization. The reason is that the pair of points at $z=0$ that
satisfy the symmetry $\SD$ also satisfy $\SU$, so they have to be accounted for in a
slightly different way. In particular, appendix~\ref{Simetria2} shows
that if the map contains $u$ pixels at $z=0$ and
$0 \le \phi < \pi$, and $k$ pixels at $z>0$ (note that $k+u=n/2$), then the blocks
$\Y_{eo}$ and $\Y_{oe}$ have $k/2$ rows each, and the blocks $\Y_{ee}$
and $\Y_{oo}$ have $k/2+u$ rows each (note that if $u=0$, the number of
rows of each block defaults to $n/4$). Therefore, the rank of the matrix will be
limited by the following expression ($R_2$):
\begin{align}
\nonumber
\rank (\Y) & \le \min (N_{eo}, k/2) + \min (N_{ee}, k/2+u) \\
& + \min (N_{oo}, k/2+u) + \min (N_{oe}, k/2),
\end{align}
where $N_{eo}$ is the number of spherical harmonics of even $\ell$ and
odd $m$, $N_{ee}$ is the number of spherical harmonics of even $\ell$
and even $m$, and so on.

\subsection{\texorpdfstring{\boldmath $\ST$ symmetry: $\phi \rightarrow \phi+\pi/2$}{SIII symmetry}}
\label{sec:s3}

Let us assume that we have a pixelization that satisfies the symmetry
$\SU$ and that for each pixel at $(\theta, \phi)$  there is a
corresponding pixel at $(\theta, \phi+\pi/2)$. Note that, under these
conditions, $\SD$ is also fulfilled. Under the transformation $\phi \rightarrow \phi+\pi/2$ we have:
\begin{equation} \label{RelacionM2_Texto}
Y_{\ell m}\left(\theta, \phi+\pi/2\right) =   i^{m} Y_{\ell m}\left(\theta, \phi\right),
\end{equation}
\noindent
which can be used to transform each non-null block in eq.~\eqref{BloquesY2} into a new block-diagonal structure (see appendix~\ref{Simetria3}). For example, the block $\Y_{eo}$ can be transformed:
\begin{equation}
\label{BloqueYpi}
\Y_{eo} \rightarrow
\left(\begin{array}{cc} \Y_{e1} & \0 \\ \0 & \Y_{e3} \end{array}\right).
\end{equation}
\noindent
At the end of this process we have decomposed matrix $\Y$ into a matrix with eight blocks in the diagonal:
\begin{equation}
\{ \Y_{e1}, \Y_{e3}, \Y_{e0}, \Y_{e2}, \Y_{o1}, \Y_{o3}, \Y_{o0}, \Y_{o2} \}.
\end{equation}
\noindent
The corresponding rank is limited by the expression ($R_3$): 
\begin{align}
\label{Formula4A}
\rank(\Y) & \le \min (N_{e1}, k/4) + \min (N_{e3}, k/4) \nonumber \\
& + \min (N_{e0}, k/4+u/2) + \min (N_{e2}, k/4+u/2) \nonumber \\
& + \min (N_{o1}, k/4+u/2) + \min (N_{o3}, k/4+u/2) \nonumber \\
& + \min (N_{o0}, k/4) + \min (N_{o2}, k/4),
\end{align}
\noindent
where $k$ and $u$ have the same meaning as in section~\ref{sec:s2}.
$N_{e q}$ is the number of spherical harmonics of even $\ell$ and
$q = \mathrm{mod}(m,4)$.\footnote{$\mathrm{mod}(i,j) = i-j \times \mathrm{floor}(i/j)$ denotes the modulo of $i$ and
$j$. In our case, $q=\mathrm{mod}(m,4)$ can take
four different values corresponding to  $\{ 0,1,2,3 \}$. Note that for
positive $m$, this quantity is simply given by the rest of $m$ divided by 4.} $N_{o q}$ is the number of spherical
harmonics of odd $\ell$ and $q = \mathrm{mod}(m,4)$. Note that he sum of the eight $N_{pq}$ adds up the total number of
spherical harmonics:
\begin{equation}
\sum_{q=0}^3 N_{eq} + \sum_{q=0}^3 N_{oq}  = \sum_{\ell=2}^{\ell_{\max}} 2 \ell+1 .
\end{equation}
Similarly, the sum of the pixels involved in the eight terms of
eq.~\eqref{Formula4A} adds up to $2k+2u$, i.e.\ the total number
of pixels $n$.

\section{Results for different pixelization schemes}
\label{sec:Pixelizaciones}

The particular way in which the sphere is pixelised gives rise to different symmetries and sets of pairs $(\theta, \phi)$, and therefore will have an impact on the rank of the covariance matrix. In this section we will compare the theoretical ranks imposed by the constraints introduced in the previous sections with those obtained for different pixelization schemes. In particular, we will consider five different pixelizations that have been used to analyse CMB data: \texttt{Cube} \cite{Cube2}, \texttt{Icosahedron} \cite{Icosahedron}, \texttt{Igloo} \cite{Igloo}, \texttt{GLESP} \cite{glesp1} and \texttt{HEALPix} \cite{healpix}.

However, before considering these particular pixelization schemes, let us study the case of a generic pixelization that successively fulfils symmetries $\SU$, $\SD$ and $\ST$. We can easily calculate the maximum rank for the covariance matrix for a certain maximum multipole $\lm$ given by the corresponding constraints. Table~\ref{Tabla:RangosTeoricos} shows the theoretical maximum ranks achieved by the covariance matrix for different values of $N_{\mathrm{pix}}$ and $\lm$, depending on which symmetries are satisfied. In particular, for all cases, the minimum $\lm$ necessary to achieve a non-singular matrix (i.e., whose rank is at least as large as $N_{\mathrm{pix}}$) is given. To allow for an easier comparison, we have chosen values for $N_{\mathrm{pix}}$ equal to the number of pixels of the first resolutions of the \texttt{HEALPix} and \texttt{Igloo} pixelizations (see sections~\ref{sec:healpix} and~\ref{sec:igloo} for details). In addition, the constraints imposed by the $\SD$ and $\ST$ symmetries also depend on the particular number of pixels at $z=0$. For the sake of simplicity, we have decided this generic pixelization to have the same number of pixels at $z=0$ as \texttt{HEALPix}. As one would expect, including symmetries in the pixelization implies, at least in certain cases, that a higher $\lm$ is needed in order to achieve a regular covariance matrix.

\begin{table}
\centering
\scriptsize{
\begin{tabular}{lcccccccccccccccccccc} 
\hline
$N_{\mathrm{pix}}$ & \multicolumn{3}{c}{12}	& & \multicolumn{3}{c}{48} & & \multicolumn{3}{c}{192} & & \multicolumn{3}{c}{768} \\ \hline
$\lm$ 		& 2 & 3 & 4 	 	& & 5  & 6  & 7	 & & 12 & 13 & 14	 & & 26  & 27 & 28	 \\ \hline
Cons. \\ \noalign{\hrule height 0.25pt}
$R_0$	& 5 & 12 & 12 		& & 32 & 45 & 48 & & 165 & 192 & 192 & & 725 & 768 & 768 \\
$R_1$ 	& 5 & 11 & 12 		& & 32 & 42 & 48 & & 165 & 186 & 192 & & 725 & 761 & 768 \\
$R_2$	& 5 & 11 & 12 		& & 32 & 42 & 48 & & 165 & 186 & 192 & & 725 & 761 & 768 \\
$R_3$  & 5 & 11 & 12 		& & 32 & 42 & 48 & & 165 & 186 & 192 & & 725 & 761 & 768 \\ \hline \hline
$N_{\mathrm{pix}}$ & & &\multicolumn{4}{c}{3072} & & & \multicolumn{3}{c}{12288}	& & \multicolumn{3}{c}{49152} \\ \hline
$\lm$ 		& & &  53 & 54 & 55 & 56 & 	 & & 109   & 110   & 111 	& & 220  & 221  & 222 \\ \hline
Cons. \\ \noalign{\hrule height 0.25pt}
$R_0$	& & & 2912 & 3021 & 3072 & 3072 &  & & 12096 & 12288 & 12288 & & 48837 & 49152 & 49152 \\
$R_1$ 	& & & 2912 & 3018 & 3072 & 3072 &  & & 12096 & 12246 & 12288 & & 48837 & 49106 & 49152 \\
$R_2$	& & & 2912 & 3017 & 3071 & 3072 &  & & 12096 & 12246 & 12288 & & 48837 & 49106 & 49152 \\
$R_3$  & & & 2912 & 3017 & 3071 & 3072 &  & & 12096 & 12246 & 12288 & & 48837 & 49106 & 49152 \\ \hline
\end{tabular}
}
\caption{Theoretical maximum ranks $R_{\mathrm{th}}$ of the covariance matrix, for a generic
pixelization of the sphere, imposed by the constraints given in
the previous section. Different number of pixels ($N_{\mathrm{pix}}$) and
maximum multipole ($\lm$) are considered.}
\label{Tabla:RangosTeoricos}
\end{table}

Although the ranks given in table~\ref{Tabla:RangosTeoricos} correspond to the maximum ranks that the covariance matrix can achieve under the presence of the corresponding symmetries, in practice, a specific pixelization may have additional properties or cancellations. This reduces even further the rank for a given multipole, which implies that we need a higher $\lm$ for this matrix to become regular. In the following subsections we will study the actual rank of the covariance matrix for five specific pixelizations, considering different resolutions. To carry out this analysis, we have calculated numerically the rank of $\C$ using the \texttt{MatrixRank} function of the symbolic computation software \textsc{Mathematica}. Hereinafter we will refer to this value as $R_{N}$. For comparison, this value is confronted with that obtained from the theoretical constraints, taking into account the symmetries present in each of the considered pixelizations, which we will denote $R_{\mathrm{th}}$.

It is worth noting that the particular elements of the covariance matrix depend on the power spectrum of the fluctuations but, according to eq.~\eqref{FinalDesarrollo}, its rank does not (except in the null case of a blank image).\footnote{Note also that, in practice, very small values of the power spectrum can introduce numerical errors in the calculation of the rank of $\C$, giving rise to ill-conditioned matrices in cases expected to be non-singular from theoretical arguments (see section~\ref{sec:Espectro} for details).} This allows us to choose the $C_{\ell}$'s and, for the sake of simplicity, in this section we will compute $\C$ for a flat spectrum $C_{\ell}=1$. For this simplified case, making use of the addition theorem for spherical harmonics, eq.~\eqref{Sumatorio} takes the form:
\begin{equation}
\label{CFormaHabitual}
\C_{ij} = \sum_{\ell=2}^{\ell_{\max}} \dfrac{2 \ell + 1}{4 \pi} \mP_{\ell}(\vvec{r}_i \vvec{r}_j) = \sum_{\ell=2}^{\ell_{\max}}  \mP^{\ell}(\vvec{r}_i \vvec{r}_j).
\end{equation}
As expected, in all the considered cases we have found $R_{N}  \equiv  \rank(\C) =\rank(\Y)$.

\subsection{\texttt{Cube}}
\label{sec:cube}

\begin{figure}
\begin{center}
\includegraphics{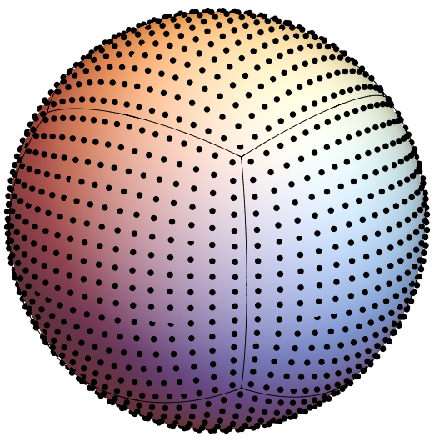}
\caption{\texttt{Cube} pixelization corresponding to  $\pr=5$ and $N_{\mathrm{pix}}=1536$.}
\label{FigCube}
\end{center}
\end{figure}

The \texttt{Cube} pixelization \cite{Cube2}, represented in figure~\ref{FigCube}, maps the pixels from the surface of a cube to the unit sphere in such a way that their areas are approximately equal. The successive resolution levels are attained by recursive subdivision of the projected faces of the cube, controlled by the resolution parameter $\pr$, which can take positive integer values. The number of pixels is given by $N_{\mathrm{pix}} = 6 \times 4^{\pr-1}$. The minimum resolution, six pixels at the centres of the six faces of the cube, is given by $\pr=1$. This pixelization has been extensively used during the processing of the COBE experiment \cite{ben96}.

In this pixelization all pixels have a $\SU$-symmetric pixel. They also satisfy the  $\ST$ symmetry, except for two pixels on the poles for the first resolution level. Table~\ref{TablaRangosCube} shows the values of the rank of the covariance matrix obtained numerically ($R_N$) and the maximum theoretical rank ($R_{\mathrm{th}}$) as a function of $\lm$. The pixels of resolutions 2 and 3 satisfy the $\ST$ symmetry, thus the maximum theoretical rank is given by $R_3$. For $\pr=1$, the theoretical ranks have been calculated taken into account that the $\SU$ symmetry is satisfied and that there are two pairs of pixels that also satisfy $\ST$.

\begin{table}
\begin{center}
\begin{tabular}{lcccccccccccccccccccc} \hline 
($\pr$, $N_{\mathrm{pix}}$) 	 	& \multicolumn{3}{c}{(1,6)} & & \multicolumn{4}{c}{(2, 24)} & & \multicolumn{10}{c}{(3,96)} \\ \hline
$\lm$ 	  				& 2 & 3 & 4 &  & 2 & 3 & 4 & 5  	  	 	& & 2 & 3 & 4 & 5 & 6 & 7 & 8 & 9 & 10 & 11 \\ \hline
$R_N$ 	  				& 2 & 5 & 6 &  & 5 & 12 & 19 & 24 			& & 5 & 12 & 21 & 32 & 45 & 60 & 75 & 88 & 94 & 96  \\
$R_{\mathrm{th}}$  	& 3 & 6 & 6 &  & 5 & 12 & 19 & 24			& & 5 & 12 & 21 & 32 & 45 & 60 & 77 & 92 & 96 & 96  \\ \hline 
\end{tabular}
\caption{Numerically calculated rank for the covariance matrix ($R_N$)
compared to the maximum expected theoretical rank ($R_{\mathrm{th}}$) for different configurations
of the \texttt{Cube} pixelization.}
\label{TablaRangosCube}
\end{center}
\end{table}
The table shows that for some $\lm$ we have $R_{N} \le
R_{\mathrm{th}}$.  In order to understand why the maximum range is not
always achieved, let us consider as a workable example the case
$\pr=1$, $\ell_{\max}=2$, where we find $R_N=2$ versus
$R_{\mathrm{th}}=3$. For this resolution level we have six pixels
located at $( \pm 1,0,0), (0,\pm 1,0), (0,0,\pm 1)$ in Cartesian
coordinates. Since $\C$ is calculated using the harmonics of $\ell =
2$, $\Y$ is a matrix with 6 rows (number of pixels) and 5 columns
(number of harmonics). Following the arguments of the previous
section, it is convenient to order the spherical harmonics following
the sequence $\{e1, e3, e0, e2\}$ (note that odd values of $\ell$ are
not considered since $\ell_{\min}=\ell_{\max}=2$). For our case, this corresponds
to an ordering of $m$ of $\{1, -1, 0, -2, 2\}$. Pixels are ordered (0,
0, 1), (1, 0, 0), (0, 1, 0), (-1, 0, 0), (0, -1, 0), (0, 0, -1). In
this way, carrying out the corresponding operations, we get the matrix
$\Y$:
\begin{equation}
\Y = 	\left(
\begin{array}{ccccc}
0 & 0 & \frac{\sqrt{\frac{5}{\pi }}}{2} & 0 & 0 \\
0 & 0 & -\frac{\sqrt{\frac{5}{\pi }}}{4} & \frac{1}{4} \sqrt{\frac{15}{2 \pi }} & \frac{1}{4} \sqrt{\frac{15}{2 \pi }} \\
0 & 0 & -\frac{\sqrt{\frac{5}{\pi }}}{4} & -\frac{1}{4} \sqrt{\frac{15}{2 \pi }} & -\frac{1}{4} \sqrt{\frac{15}{2 \pi }} \\
0 & 0 & -\frac{\sqrt{\frac{5}{\pi }}}{4} & \frac{1}{4} \sqrt{\frac{15}{2 \pi }} & \frac{1}{4} \sqrt{\frac{15}{2 \pi }} \\
0 & 0 & -\frac{\sqrt{\frac{5}{\pi }}}{4} & -\frac{1}{4} \sqrt{\frac{15}{2 \pi }} & -\frac{1}{4} \sqrt{\frac{15}{2 \pi }} \\
0 & 0 & \frac{\sqrt{\frac{5}{\pi }}}{2} & 0 & 0 \\
\end{array}
\right).
\end{equation}
The first two columns of zeros occur because $Y_{2}^{1}$ and
$Y_{2}^{-1}$ contain the product $\sin{\theta}\cos{\theta}$, which
becomes null for the 6 considered pixels. All points are symmetric in the sense $\vvec{r} \rightarrow -\vvec{r}$, and therefore we can use eq.~\eqref{SimetriaY_Texto} and transform $\Y$ into a matrix with 3 rows of zeros. Choosing the pixels $ (0,0,1)$, $ (1,0,0)$ and $ (0,1,0)$ when applying the symmetry we get:
\begin{equation}
\Y \rightarrow	\left(\begin{array}{ccccc}
0 & 0 & \frac{\sqrt{\frac{5}{\pi }}}{2} & 0 & 0 \\
0 & 0 & -\frac{\sqrt{\frac{5}{\pi }}}{4} & \frac{1}{4} \sqrt{\frac{15}{2 \pi }} & \frac{1}{4} \sqrt{\frac{15}{2 \pi }} \\
0 & 0 & -\frac{\sqrt{\frac{5}{\pi }}}{4} & -\frac{1}{4} \sqrt{\frac{15}{2 \pi }} & -\frac{1}{4} \sqrt{\frac{15}{2 \pi }} \\
0 & 0 & 0 & 0 & 0 \\
0 & 0 & 0 & 0 & 0 \\
0 & 0 & 0 & 0 & 0 \\
\end{array}
\right).
\end{equation}
Then, the maximum rank we could ever achieve is 3. Note that
$Y_{2}^{2} = Y_{2}^{-2} = 0$ at $(0, 0, 1)$, so the first row in
$\Y_{e}$ has only one non zero term, at the column that corresponds to
$Y_{2}^{0}$. Taking into account that the other two points are $\ST$
symmetric, we can get more zero blocks in $\Y_e$. Therefore, the first
three rows of the matrix become:
\begin{equation}
\Y_e \rightarrow	\left(
\begin{array}{ccccc}
0 & 0 & \frac{\sqrt{\frac{5}{\pi }}}{2} & 0 & 0 \\
0 & 0 & -\frac{\sqrt{\frac{5}{\pi }}}{4} & 0 & 0 \\
0 & 0 & 0 & -\frac{1}{4} \sqrt{\frac{15}{2 \pi }} & -\frac{1}{4} \sqrt{\frac{15}{2 \pi }} \\
\end{array}
\right),
\end{equation}
which yields to a matrix of rank 2, a unit of rank lower than expected
taking into account the size of $Y_e$ after applying $\SU$. The reason
for this lower rank for this particular pixelization is that two columns of spherical harmonics, $Y_{2}^{-1}$ and $Y_{2}^{1}$, are zeros and the columns of $Y_{2}^{2}$ and $Y_{2}^{-2}$ are equal. This illustrates how in some cases the rank can be lower than expected by the theoretical expressions when computing the rank of matrices $\Y$ calculated on particular sets of pixels.

\subsection{\texttt{Icosahedron}}

\begin{figure}
\begin{center}
\includegraphics{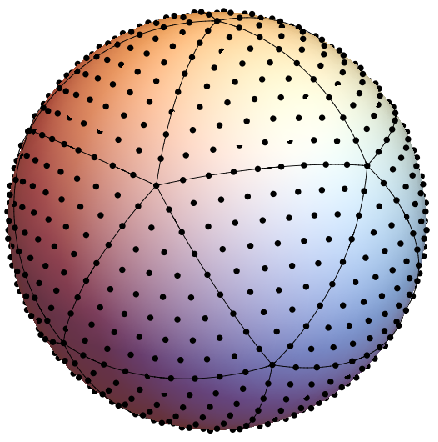}
\caption{\texttt{Icosahedron} pixelization corresponding to $\pr=5$ and $N_{\mathrm{pix}}$=812.}
\label{FiguraIcosahedron}
\end{center}
\end{figure}

The \texttt{Icosahedron} pixelization \cite{Icosahedron}, shown in
figure~\ref{FiguraIcosahedron}, is constructed by subdividing the
faces of an icosahedron into a regular triangular grid, starting from a 
first resolution of 12 pixels situated at the vertices of the icosahedron. The pixels
are projected onto the sphere in such a way that their areas are approximately equal.
Similarly to the \texttt{Cube} pixelization, the resolution of the pixelization
is controlled by an integer positive parameter $\pr$, being the
number of pixels given by $N_{\mathrm{pix}} = 40 \times{\pr} \times{(\pr-1)}+12$.

This pixelization satisfies the $\SU$ symmetry. However, since the vertices of the Icosahedron, which are the starting point of the subsequent resolutions, do not fulfil symmetries $\SD$ or $\ST$, no further symmetries are expected to be found. In particular, we have tested that this is the case up to resolution $\pr=9$. Table~\ref{TablaRangosIcosahedron} shows the values of the rank of the covariance matrix obtained numerically ($R_N$) and the maximum theoretically expected range ($R_{\mathrm{th}}$) as a function of $\lm$ for two different resolutions. Note that for the shown resolutions the $\SU$ symmetry is satisfied and, therefore, $R_{\mathrm{th}} = R_1$.
\begin{table}		
\begin{center}
\begin{tabular}{lccccccccccccccccc} \hline
($\pr$, $N_{\mathrm{pix}}$) & \multicolumn{5}{c}{(1, 12)} & & \multicolumn{10}{c}{(2, 92)} \\ \hline
$\lm$ 		& 2 & 3 & 4 & 5 & 6 		 & & 3 & 4 & 5 & 6 & 7 & 8 & 9 & 10 & 11 & 12 \\ \hline
$R_N$ 		& 5 & 8 & 8 & 11 & 12		 & & 12 & 21 & 32 & 45 & 60 & 77 & 87 & 88 & 91 & 92\\
$R_{\mathrm{th}}$ 	& 5 & 11 & 12 & 12 & 12		 & & 12 & 21 & 32 & 45 & 60 & 77 & 90 & 92 & 92 & 92\\ \hline
\end{tabular}
\caption{Numerically calculated rank for the covariance matrix ($R_N$)
compared to the maximum expected theoretical rank ($R_{\mathrm{th}}$) for different configurations
of the \texttt{Icosahedron} pixelization.}
\label{TablaRangosIcosahedron}
\end{center}
\end{table}
As for the \texttt{Cube} pixelization, $R_N \leq R_{\mathrm{th}}$ is always satisfied. The reason why the maximum rank is not always achieved can be easily seen in the lowest resolution ($\pr=1$, $N_{\mathrm{pix}} = 12$). If we sum up to $\lm=2$, the rank of the matrix is limited by the number of spherical harmonics, which is 5, and we have found $R_N=5$. If we sum up to $\lm=3$,  there are 7 new harmonics, thus the rank could increase, in principle, up to 12. However, due to the presence of the $\SU$ symmetry, we have that the maximum theoretical rank is $R_{\mathrm{th}}=11$. Nevertheless, the numerical rank $R_N$ is found to be 8. One of the reasons for the reduction of the rank is that the z-coordinate of all pixels take one of the values $z=\{-1,1,-\frac{1}{\sqrt{5}},\frac{1}{\sqrt{5}}\}$ and the Legendre polynomials for $\ell=3$ and  $m =\pm 1$,
\begin{equation}
P_3^{-1}(z) = \frac{1}{8} \sqrt{1-z^2} \left(5 z^2-1\right), \qquad
P_3^{1}(z)  = -\frac{3}{2} \sqrt{1-z^2} \left(5 z^2-1\right),
\end{equation}
\noindent
happen to cancel at these points, which leads to two zero columns in the $\Y_o$ block. Thus we are left with only five non-zero columns instead of seven for this block, what means that the covariance matrix rank could be ten as maximum. However, due to the particular positions of the pixels in this pixelization scheme, other columns of this block happen to be linearly dependent. In particular,  the column that corresponds to $m=0$ is linearly independent of the other four columns, but the $m=-3$ column is proportional to the $m=-2$ column and the same relation applies to columns $m=3$ and $m=2$. Therefore, the rank of the odd block is reduced to 3, and adding up the rank of the even block, which is 5, we finally get rank 8 for the covariance matrix as found.  Note that adding multipoles up to $\lm=4$ does not change the situation, but going up to $\lm=5$ raises the rank to 11. Finally, adding multipoles up to $\lm=6$ provides enough independent columns to get a non-singular covariance matrix.

\subsection{\texttt{Igloo}}
\label{sec:igloo}
The \texttt{Igloo} pixelation \cite{Igloo} divides the sphere in bands which are perpendicular to the axis $z$ in variable intervals of $\theta$. Each band is divided into a different number of pixels at constant intervals of $\phi$, and the interval varies from band to band. Thus, the edges of the pixels are of constant latitude and longitude. Among the different options, the one presented here consists of a base resolution of twelve pixels, three on each of the poles and six on the central band. The angles have been chosen in such a way that the resultant pixels are of equal area. The subsequent resolutions were obtained by recurrent subdivisions of each pixel into other four pixels of equal area, always having three pixels at each pole.

\begin{figure}
\begin{center}
\includegraphics{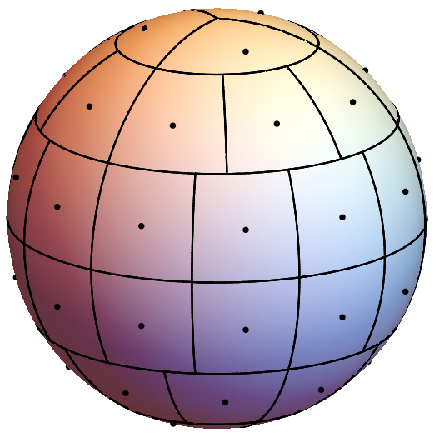}
\caption{\texttt{Igloo} pixelization corresponding to $N_{\mathrm{side}}$=2 and $N_{\mathrm{pix}}$=48.}
\label{FiguraIgloo}
\end{center}
\end{figure}

The number of pixels is $N_{\mathrm{pix}}= 12 \times N_{\mathrm{side}}^2$. The resolution is controlled by the $N_{\mathrm{side}}$ parameter, which can take as values integers that are a power of 2.  Figure~\ref{FiguraIgloo} shows the \texttt{Igloo} pixelization for $N_{\mathrm{side}}=2$. The Igloo pixels are symmetric in the $\SU$ sense. Moreover, for all resolutions there are couples of $\SD$-symmetric pixels in the central area. For resolutions 2 and higher, there are also some pixels with $\ST$ symmetry.
Taking into account these particular symmetries, we find that $R_N =  R_{\mathrm{th}} = R_1$ for all $\lm$ for resolutions $N_{\mathrm{side}}=\{1,2\}$. For $N_{\mathrm{side}} \geq 4$ the numerical rank of the covariance matrix is found to be smaller than the maximum theoretical rank in some cases. Table~\ref{TablaRangosIgloo} shows the results for $N_{\mathrm{side}}=4$ and different values of $\lm$.
\begin{table}
\begin{center}
\begin{tabular}{lccccccccccccccc} \hline
($N_{\mathrm{side}}, N_{\mathrm{pix}}$)			& \multicolumn{13}{c}{(4, 192)} &	& \\ \hline
$\lm$ 		& 2 & 3 & 4 & 5 & 6 & 7 & 8	& 9 & 10 & 11 & 12 & 13 & 14 & 15 & 16 \\ \hline 
$R_N$ 		& 5 & 12 & 21 & 32 & 45 & 60 & 77 &	96 & 117 & 140 & 162 & 179 & 187 & 191 & 192  \\
$R_{\mathrm{th}}$ 	& 5 & 12 & 21 & 32 & 45 & 60 & 77 &	96 & 117 & 140 & 165 & 186 & 192 & 192 & 192 \\ \hline
\end{tabular}
\caption{Numerically calculated rank for the covariance matrix ($R_N$)
compared to the maximum expected theoretical rank ($R_{\mathrm{th}}$) for different configurations
of the \texttt{Igloo} pixelization.}
\label{TablaRangosIgloo}
\end{center}
\end{table}

\subsection{\texttt{GLESP}}\label{sub:glesp}

\begin{figure}
\begin{center}
\includegraphics{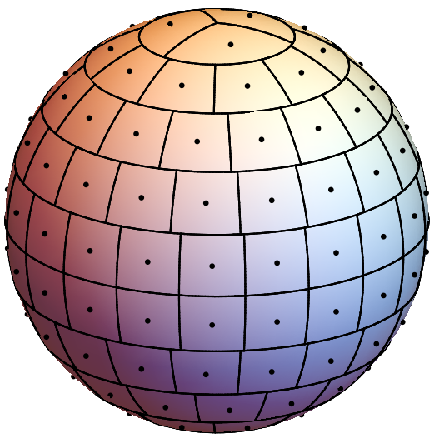}
\caption{\texttt{GLESP} pixelization corresponding
to a resolution parameter
$N=10$. In this figure the rings have been
rotated to make the pixels
$\SU$-symmetric.}
\label{FiguraGLESP}
\end{center}
\end{figure}

The Gauss-Legendre Sky Pixelization (\texttt{GLESP}) \cite{glesp1} (figure~\ref{FiguraGLESP}) makes use of the Gaussian quadratures to evaluate numerically the integral with respect to $z$ of the expression:
\begin{equation}
a_{\ell m} = \int_{-1}^{1}{dz} \int_{0}^{2 \pi} d \phi \Delta T(z,\phi) Y^{*}_{\ell m}(z,\phi) ,
\end{equation}
which can be formally expressed with this method in an exact form as a
weighted finite sum. The surface of the sphere is
divided into $N$ rings of trapezoidal pixels with values of $\theta$
at the centres of the pixels according to the Gauss-Legendre
quadrature method. There is some degree of freedom to fix the number
and size of the pixels on the rings, but preferentially they are
defined to make the equatorial ones roughly square. The number of
pixels on the rest of the rings is chosen in such a way that all the
pixels have nearly equal area.\footnote{It is interesting to note that to study the CMB
polarization and with the aim of obtaining a better evaluation of
the $_{0,\pm 2} a_{\ell, m}$-coefficients, two additional
over-pixelization versions of this scheme have been proposed \cite{glesp2}. In this
case, there is a larger number of pixels on the rings than in the
nearly equal area scheme considered in this paper, with decreasing area of
the pixels on the rings as they get closer to the polar caps. Since our work is oriented to the scalar field case,
we will study the nearly equal area version of the pixelization.} Finally, polar pixels are
triangular. Note that this pixelization is not
hierarchical.

Regarding the presence of symmetries, those pixels belonging to rings which have been divided into a number of pixels which is a multiple of two or four will satisfy $\SD$ or $\ST$. However, this is not always the case and, therefore, the \texttt{GLESP} pixelization does not fulfil these symmetries as a whole. Referring to the $\SU$-symmetry, it can be fulfilled in certain cases if the rings are rotated with respect to the $z$ axis (note that this rotation does not change the essence of the pixelization). In particular, if the number of rings is even, each of the rings on one hemisphere can be positioned with respect to the opposite ring on the other hemisphere in such a way that the $\SU$-symmetry is satisfied. If $N$ is an odd number, there is an unmatched ring on the equator, and the symmetry is only fulfilled if this ring is divided into an even number of pixels. Fig.~\ref{FiguraGLESP} shows the GLESP pixelization for $N=10$ in a $\SU$-symmetric configuration.

Table~\ref{TablaRangosGLESP} shows the values of the rank for the case with 4 rings. The number of pixels is 22, 8 on each of the central rings and 3 on each of the poles. For the sake of simplicity and in order to allow for a better comparison with the other considered pixelizations, we have rotated the rings in such a way that all the pixels have their $\SU$-symmetric pair. The pixels on the two central rings have a $\SU$, $\SD$ and $\ST$-symmetric partner, but the pixels on the poles do not have a $\SD$ and $\ST$-symmetric one. In this situation, the $R_{II}$ and $R_{III}$ expressions are not strictly valid. However, if we find that $R_{I}=R_{II}=R_{III}$, i.e., that the matrix rank is not reduced under the presence of $\SD$ and $\ST$, we can conclude that the same is valid when the symmetries are only partially fulfilled, as for the \texttt{GLESP} pixelization. For the considered case, we have indeed found $R_{I}=R_{II}=R_{III}$ for all the values of $\lm$. Table~\ref{TablaRangosGLESP} shows that for $\lm=4$ and $\lm=5$ the rank of $\C$ is one unit lower than the theoretical rank. Let us try to explain where this unit is lost. Since the $\SU$-symmetry is fulfilled, we have two diagonal blocks in $\Y$. For $\lm=4$, the block of odd $\ell$ has 11 rows and 7 columns, and its rank is 7. The block of even $\ell$, 11 rows and 14 columns, and its rank is 10. Due to the way in which the values of $\theta$ are chosen in this pixelization scheme, the roots of the Legendre polynomial with $\ell=4$ in the case of $N=4$, the column of $\Y$ that corresponds to the harmonic $Y_{4, 0}$ is made up of zeros. Aside from this column, among the other 13 columns there are only 10 linearly independent columns. We have found that the group of ten that corresponds to the five harmonics of values of $\ell=2$ and the other five of $\ell=3$ and values of $m:$ -4, -3, -1, 1 and 3 are linearly independent. When $\lm=5$, we get 11 new columns on the odd block of $\Y$, and the rank in this block saturates to 11, but the block of even $\ell$ still has 11 rows and 14 columns but only 10 linearly independent columns; thus the rank is a unit lower than the theoretical value. When $\lm=6$ we get new linearly independent columns in the block of even $\ell$, and the rank saturates.
\begin{table}
\begin{center}
\begin{tabular}{lccccc} \hline
($N, N_{\mathrm{pix}}$)			& \multicolumn{4}{c}{(4, 22)} 	& \\ \hline
$\lm$ 		& 2 & 3 & 4 & 5 & 6  \\ \hline
$R_N$ 		& 5 & 12 & 17 & 21 & 22  \\
$R_{\mathrm{th}}$ 	& 5 & 12 & 18 & 22 & 22  \\ \hline
\end{tabular}
\caption{Numerically calculated rank for the covariance matrix ($R_N$)
compared to the maximum expected theoretical rank
($R_{\mathrm{th}}$) for the \texttt{GLESP} pixelization for $N=4$.}
\label{TablaRangosGLESP}
\end{center}
\end{table}

\subsection{\texttt{HEALPix}}\label{sec:healpix}

\begin{figure}
\begin{center}
\includegraphics{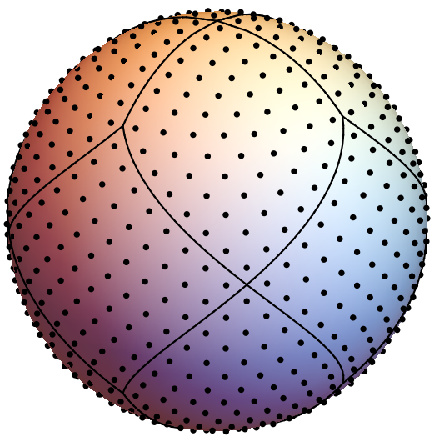}
\caption{ \texttt{HEALPix} pixelization corresponding to $N_{\mathrm{side}} =
8$ and $N_{\mathrm{pix}}$=768.}
\label{FiguraHealpix}
\end{center}
\end{figure}

The most extensively used pixelization for CMB analysis is
\texttt{HEALPix}, which stands for Hierarchical Equal Area
iso-Latitude Pixelization \cite{healpix}. \texttt{HEALPix} divides
the sphere into twelve spherical diamond-shaped pixels on three rings,
one of them on the equator and the other two towards the
poles. The four pixels forming each ring have the
same latitude and each diamond is then subdivided recursively in order to get the pixels for the
different resolution levels. The total number of pixels is $N_{\mathrm{pix}} =
12 \times N_{\mathrm{side}}^2$, where $N_{\mathrm{side}}$ is the resolution parameter,
which is always an integer power of 2. Figure~\ref{FiguraHealpix} shows
the \texttt{HEALPix} pixelization for $N_{\mathrm{side}}$=8.
By construction all the pixels satisfy the $\SU$, $\SD$ and $\ST$
symmetries. We find that for the three lowest resolutions
($N_{\mathrm{side}}=1,2$ and $4$), the covariance matrix always achieve the
maximum possible rank, i.e., $R_N = R_{\mathrm{th}} =R_3$ for all $\lm$.  The first
reduction of rank occurs for $N_{\mathrm{side}}=8$ and
$\ell_{\max}=26$ (see table~\ref{TablaRangosHealpix}). In particular, when $\lm=26$ we have $R_{\mathrm{th}}=725$
and $R_N=724$. Let us focus on this case in more detail. We have
found that the loss of rank of the covariance matrix is found to lie in block
$\Y_{e2}$. 
\begin{table}
\begin{center}
\begin{tabular}{lccccc} \hline
($N_{\mathrm{side}}, N_{\mathrm{pix}}$)			& \multicolumn{4}{c}{(8, 768)} 	& \\ \hline
$\lm$ 		& 25 & 26 & 27 & 28  \\ \hline
$R_N$ 		& 672 & 724 & 760 & 768  \\
$R_{\mathrm{th}}$ 	& 672 & 725 & 761 & 768  \\ \hline
\end{tabular}
\caption{Numerically calculated rank for the covariance matrix ($R_N$)
compared to the maximum expected theoretical rank
($R_{\mathrm{th}}$) for the \texttt{HEALPix} pixelization for $N_{\mathrm{side}}=8$
in the multipole regime in which $\C$ becomes regular.}
\label{TablaRangosHealpix}
\end{center}
\end{table}
The number of rows in block $\Y_{e2}$ is determined by the quantity
$k/4+u/2$, while the number of columns is given by the number of
spherical harmonics with even $\ell$ and $mod(m,4)=2$ (see
appendix~\ref{Simetria3} for details). In particular, for $N_{\mathrm{side}}=8$
we have a total of 768 pixels. Among these, 32 have $z=0$, one half of
them with $0 \le \phi < \pi$ (i.e. $u=16$). We also have 368 pixels
with $z>0$ (i.e. $k=368$). Therefore, the block $\Y_{e2}$ has
$k/4+u/2=100$ rows. For $\ell_{\max}=26$, the number of spherical
harmonics with $\ell$ even and {$ mod(m,4)=2$ is 98. So the block has
size $100 \times 98$, but rank 97. We have also checked that the
rank of all the sub-blocks that can be formed by removing any
particular column of $\Y_{e2}$ is 97. This means that any group of
97 out of the 98 spherical harmonics are linearly independent, but
the 98 columns taken as a whole are linearly coupled. This shows,
again, how some units of rank can be lost by properties related to
the particular values of the pixels of the chosen pixelization.

\bigskip

So far we have demonstrated, for five different pixelization schemes, that for a given resolution parameter, it is possible to determine the lowest $\lm$ needed to make $\C$ regular. Furthermore, a detailed analysis of the pixel properties allows us to detect linear combinations of spherical harmonics that lead to reductions of the maximum theoretical rank. When this happens, it is necessary to raise $\lm$ in order to regain regularity.

Among the five pixelization schemes we have considered, \texttt{HEALPix} is particularly interesting because of two nice properties. On the one hand, from the tests we have presented, it seems to be the pixelization in which the reduction of the rank of the covariance matrix with respect to the maximum expected theoretical value is lowest. On the other hand, since it presents the three types of considered symmetries for all the resolution levels, it is easy to divide the $\Y$ matrix into eight blocks, which in turn makes it easier to study the linear independence of the columns of the matrix. For these reasons, together with the fact that \texttt{HEALPix} is the most commonly used pixelization for CMB analysis, we will only consider this pixelization along the rest of the paper.

The discussion so far has focused on an idealized case. In particular, there is a number of real life issues we have not taken into account yet:

\begin{itemize}
	\item Until now we have used $C_{\ell}=1$ and neglected the window function $B_{\ell}$.
	\item In many real life applications $\C$ is calculated for a masked area of the sky.
	\item We have not considered the effect of noise.
\end{itemize}

We will discuss the impact of these effects on the regularity of $\C$ in the next sections.

\section{Effect of the power spectrum and the beam transfer function}
\label{sec:Espectro}

Up to now we have studied the rank of $\C$ according to eq.~\eqref{Sumatorio}
using a simplified angular power spectrum
$C_{\ell}=1$. In a realistic CMB analysis,
we would use the corresponding $C_{\ell}$'s and the beam transfer function
$B_{\ell}$ which encodes the response of the beam of the experiment. Therefore:
\begin{equation}
\label{CFormaHabitualBl}
\C_{ij} = \sum_{\ell}\dfrac{2 \ell+1}{4 \pi} C_{\ell}B_{\ell}^2
\mP_{\ell}(\vvec{r}_i \vvec{r}_j).
\end{equation}
The beam transfer function usually drops quickly for high multipoles and the sum
of eq.~\eqref{CFormaHabitualBl} is dominated by low-$\ell$ terms. As we
mentioned before, theoretically this should not affect the rank of
$\C$. However, in practice, since the sums are computed numerically, it is possible that
the regularity of $\C$ is affected by machine precision
limitations. In table~\ref{TablaRangosBl} we compare the rank of $\C$
as calculated in the previous section (i.e. $C_{\ell}$=1 for all
multipoles and not including any window function) with that calculated using a realistic CMB power spectrum,
for the \texttt{HEALPix} pixelization with resolution  $N_{\mathrm{side}}=8$. In
particular, we have considered the Planck $\Lambda$CDM best-fit
model that includes also information from the Planck lensing power
spectrum reconstruction and Baryonic
Acoustic Oscillations \cite{2016A&A...594A..13P},\footnote{More specifically we have used model 2.30 described in the document \url{https://wiki.cosmos.esa.int/planckpla2015/images/f/f7/Baseline_params_table_2015_limit68.pdf} from the Planck Explanatory Supplement}
a Gaussian beam with a FWHM corresponding to
2.4 times the pixel size at the considered resolution (FWHM=17.6
degrees for $N_{\mathrm{side}}=8$) and we have also
taken into account the corresponding \texttt{HEALPix} pixel window
function. This model will be used along the paper, unless otherwise
stated. As before, we have estimated the rank of the covariance
matrix using \textsc{Mathematica} with its default machine
precision (which corresponds to the standard double precision). 
From the results of table~\ref{TablaRangosBl}, we see that the rank estimated for the
covariance matrix can be significantly smaller when working under
realistic conditions for the power spectrum, indicating that
numerical errors may have an important effect in real life applications.
\begin{table}
\begin{center}
\begin{tabular}{ccccccccccc} \hline
$\lm$     & 24 & 25 & 26 & 27 & 28 & 29 & 30 & 31 \\ \hline
$R_N(1)$   & 621 & 672 & 724 & 760 & 768 & 768 & 768 & 768 \\
$R_N(2)$   & 621 & 671 & 709 & 735 & 752 & 761 & 766 & 768 \\ \hline
\end{tabular}
\caption{Numerically calculated ranks of $\C$ for a \texttt{HEALPix} resolution of
$N_{\mathrm{side}}=8$ and different values of $\lm$. $R_N(1)$ is the rank of $\C$ using $C_{\ell}=1$ and
not including any beam window function. $R_N(2)$ is the rank of the
same matrix when considering a realistic CMB power spectrum and beam.}
\label{TablaRangosBl}
\end{center}
\end{table}
We can further study the effect of numerical precision on the
effective rank of $\C$ by means of the \textsc{Mathematica}
software. \textsc{Mathematica} can operate the sums of spherical
harmonics symbolically and convert the result into numerical format
with any desired number of decimals. Those quantities which are
available with a limited
precision (e.g. the power spectrum or the pixel window function) can
be rationalized in \textsc{Mathematica}, which helps to control the
propagation of numerical errors.\footnote{Of course this comes at the cost of increasing the
computational time.} In this way, calculating the covariance matrix
given by eq.~\eqref{CFormaHabitualBl} with a precision of 200 decimals we recover
the ranks given by $R_N(1)$ in table~\ref{TablaRangosBl}. This
confirms that the loss of rank observed in table~\ref{TablaRangosBl} is in fact a matter of
numerical errors.

The effect of numerical precision can be observed in more detail in table~\ref{TablaDeterminantesPrecisiones}. We have calculated the determinant of $\C$ for a realistic power spectrum for different values of $\lm$ and different numerical precisions: the machine native precision and 50, 200, 400 and 800 decimals. For $\lm=26$ and $\lm=27$ the determinant must be zero because the theoretical rank is lower than 768. This fact is easily seen at higher numerical precisions, where the value of the determinant quickly drops  to zero. Conversely, at $\lm=$ 28, 29 and 30, the determinant is non-zero, and, even if very small, we see that its value is stable when sufficiently high precision is used. In particular, the default machine precision is not enough to get the correct value of the determinant (considering a precision of two decimal places), although it gets closer to the correct value as $\lm$ increases. This relates to our previous finding showing that the estimated rank is lower than expected for these values of $\lm$ when using the default machine precision, and it becomes correct for higher numerical precisions.

\begin{table}
\begin{center}
\small{
\begin{tabular}{cccccc} \hline
$\ell_{\max}$ & Machine & 50 &  200 & 400 & 800 \\ \hline
26  & $ -7.70\times 10^{-615} $ & $ 2.95\times 10^{-2713} $ &  $ -2.49\times 10^{-9329} $ & $ 1.44\times 10^{-18046} $ & $ 1.05\times 10^{-35734}$ \\
27  & $ -9.50\times 10^{-390} $ & $ -4.58\times 10^{-759} $ &  $ 4.90\times 10^{-1979} $ & $ -1.74\times 10^{-3545} $ & $ -4.80\times 10^{-6767}$ \\
28  & $ 6.91\times 10^{-255} $ & $ 3.12\times 10^{-256} $ & $ 3.12\times 10^{-256} $ & $ 3.12\times 10^{-256} $ & $ 3.12\times 10^{-256}$ \\
29  & $ 1.64\times 10^{-174} $ & $ 2.62\times 10^{-174} $ & $ 2.62\times 10^{-174} $ & $ 2.62\times 10^{-174} $ & $ 2.62\times 10^{-174}$ \\
30  & $ 5.54\times 10^{-136} $ & $ 5.66\times 10^{-136} $ & $ 5.66\times 10^{-136} $ & $ 5.66\times 10^{-136} $ & $ 5.66\times 10^{-136}$ \\
31  & $ 4.39\times 10^{-122} $ & $ 4.40\times 10^{-122} $ & $ 4.40\times 10^{-122} $ & $ 4.40\times 10^{-122} $ & $ 4.40\times 10^{-122}$ \\ \hline
\end{tabular}}
\end{center}
\caption{Determinant of $\C$ with different precisions and $\lm$.}
\label{TablaDeterminantesPrecisiones}
\end{table}

\section{Effect of noise}
\label{sec:Ruido}

Another common complexity on CMB data is the presence of instrumental
noise. Let us now consider the effect of adding a zero mean noise
$\vvec{n}$ to the signal $\vvec{s}$. Given that noise and signal
are expected to be statistically independent, the covariance matrix of
the data $\x = \vvec{s} + \vvec{n}$ is simply given by the sum of
the signal ($\mS$) and noise ($\N$) covariance
matrices.  Moreover, if the noise is not spatially correlated, the matrix $\N$ is diagonal. In that case: 
\begin{equation}
\label{MCRuido2}
\C_{ij}  =  \mS_{ij} + \N_{ij}
= \sum_{\ell=\ell_{\min}}^{\ell_{\max}}\dfrac{2 \ell + 1}{4 \pi}  C_\ell B_{\ell}^2 \mP_{\ell}(\vvec{r}_i \vvec{r}_j)
+ \sigma^2_i \delta_{ij}.
\end{equation}
Taking into account that each column of $\C$ is given by the sum of two columns, we can expand by columns its determinant. In this way, the expansion takes the form of the sum of $2^n$ determinants, one of them containing only columns from $\mS$, a second one containing only columns from $\N$ and the rest containing columns from both matrices. The determinants that contain columns from $\mS$ can be either positive or null, depending on the value of $\lm$ and the pixel configuration, but the term that depends only on columns from $\N$ is always positive. It can also be easily shown that the terms with columns from both matrices are also greater or equal than zero. Therefore, we have:
\begin{equation}
\det{\C} \geq \det{\N} > 0.
\end{equation}
This shows that the presence of noise regularizes the covariance matrix. This is a well known fact, and in many applications it is common to use a small amount of artificial noise to help regularizing a numerically ill-behaved covariance matrix~\cite{Planck2015_XI, 2016ApJ...825...66B}. The question arising in these cases is what is the optimal noise level to be introduced in order to make the matrix regular without degrading too much the quality of the data.

We have carried out some tests in order to illustrate the regularizing effect of adding noise to the covariance matrix for four different resolutions ($N_{\mathrm{side}}=$ 4, 8, 16 and 32) of the \texttt{HEALPix} pixelization. We have used the same Planck $\Lambda$CDM model as in the previous section for the CMB power spectrum and a Gaussian beam with FWHM equal to 2.4 times the pixel size at each resolution. The pixel window function given by \texttt{HEALPix} has also been taken into account. For each case we have performed the summations up to the $\lm$ value for which $\C$ is regular according to the constraint $R_3$. We have also added different levels of isotropic white noise according to $\sigma^2 = 10^{-f_n}$ $(\mu$K$)^2$, to be compared with the CMB variance given in the fourth column of table~\ref{RangosRuidoExperimental}. As in the previous section, the rank of the covariance matrix for each of the considered cases has been obtained using \textsc{Mathematica} with its default machine precision and is given in table~\ref{RangosRuidoExperimental}.

\begin{table}
\begin{center}
\begin{tabular}{cccccccccccc} \hline 
$N_{\mathrm{side}}$ & $N_{\mathrm{pix}}$ & $l_{\max}$ & $\mS_{ii}$  &  No noise & $f_n$ = 9.5 & 9.0 & 8.5 & 8.0 & 7.5 & 7.0 \\ \hline
4  &   192 &  14 &  487 & {\bf 192} & 192 & 192 & 192 & 192 & 192 & 192 \\
8  &   768 &  28 & 1057 & 752 &  752 & 756 &  {\bf  768} & 768 & 768 & 768\\
16 &  3072 &  56 & 1655 & 2937 & 2937 & 2937 & 2938 & {\bf 3072} & 3072 & 3072 \\
32 & 12288 & 111 & 2308 & 11518 & 11518 & 11518 & 11518 & 11518  & {\bf 12288} & 12288 \\ \hline
\end{tabular}
\caption{Rank ($R_N$) of $\C$ for several \texttt{HEALPix} resolutions after
adding different levels of noise. The columns show, from left to
right: the resolution parameter, the corresponding number of
pixels, the $\lm$ used for the calculation of $\C$ (which is the theoretical multipole that makes the matrix regular), the value of the
diagonal element $\mS_{i i}$ (i.e.\ the CMB
variance, in units of $(\mu$K$)^2$), the matrix rank without noise and the one obtained after adding noise with different values of
the  $f_n$ parameter such that
$\sigma^2 = 10^{-f_n}$. The bold face indicates the level of noise at which
the matrix $\C$ becomes regular. }
\label{RangosRuidoExperimental}
\end{center}
\end{table}

We find that for $N_{\mathrm{side}} = 4$, the covariance matrix is regular even without the presence of noise, whereas for higher resolutions it becomes necessary to add a certain level of noise for the matrix to be regular. Note that the required level of noise grows with $N_{\mathrm{side}}$, although it is always very small in comparison to the value of the CMB variance and, therefore, it is not expected to compromise the quality of the data.

In many scientific applications, as for example the estimation of the power spectrum through the Quadratic Maximum Likelihood method \cite{QML}, it is necessary to perform operations with $\C$ or its inverse. Such operations propagate numerical errors and may lead to instabilities even if one starts with a regular matrix. In figure~\ref{GraficaDeterminanteRangos} we show the result of the operation $-\log \left( {\rm abs} \left( 1-|\C \C^{-1}| \right) \right)$, with $\C$ calculated using the theoretical $\lm$ value that makes the matrix regular (third column in table~\ref{RangosRuidoExperimental}), as a function of the noise level. Positive values of this quantity indicate approximately the decimal place at which the determinant $|\C \C^{-1}|$ departs from unity, while negative values show how many orders of magnitude this determinant (in absolute value) is greater than unity. This gives an indication of the level of noise required to operate with the covariance matrix and its inverse within the required precision.
\begin{figure}
\begin{center}
\includegraphics{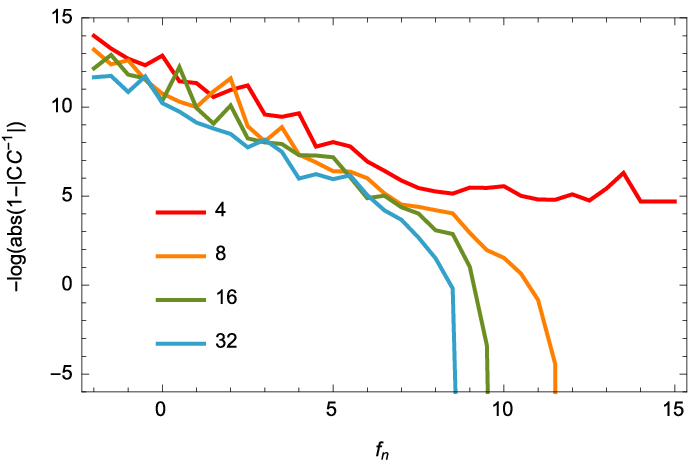}
\caption{Measurement of the departure of the numerical value of $|\C
\C^{-1}|$ from unity versus the level of added noise, for $N_{\mathrm{side}}$=
4, 8, 16 and 32 (see text for details).}
\label{GraficaDeterminanteRangos}
\end{center}
\end{figure}

\section{Effect of masking}\label{sec:Mascara}

The first effect of masking is a reduction of the number of pixels over which the covariance matrix is computed, and therefore a smaller dimension of $\C$. Consequently, a lower value of $\ell_{\max}$ will be needed in general to obtain a regular matrix. A second more subtle effect is the possible breaking of pixel -- spherical harmonic symmetries. In section~\ref{sec:Simetrias} we studied the effect of three kind of symmetries, $\SU$, $\SD$ and $\ST$. We showed that, among them, $\SU$ has the greatest impact on the rank of the covariance matrix. Therefore, for the sake of simplicity, in this section we will study the effect of masking only under the presence of the $\SU$ symmetry.

Let us consider that, for a fixed resolution parameter and a given mask geometry, we have $n$ valid pixels. Let us also consider that, among these pixels, $n_p$ have their symmetric pixel inside the region allowed by the mask, while $n_u$ pixels do not have it. Taking into account the symmetries of the $n_p$ pixels the matrix $\Y$ can be transformed as described in section~\ref{sec:Simetrias}:
\begin{equation}
\label{MatrizMascara}
\Y \rightarrow	\left( 
\begin{array}{cc}
{\Y}_{n_p/2 \times N_e}^{even} 
& 
\0_{n_p/2 \times N_o} 
\\ \\
\0_{n_p/2 \times N_e} & {\Y}_{n_p/2 \times N_o}^{odd} 
\\ \\
{\Y}_{n_u \times N_e}^{even}  &  {\Y}_{n_u \times N_o}^{odd}
\end{array}
\right),
\end{equation}
\noindent
where, for example, ${\Y}_{n_p/2 \times N_e}^{even}$ stands for
a block with $n_p/2$ rows and $N_e$ columns, whose elements are the
spherical harmonics of even $\ell$ calculated over the independent half
of symmetrical pixels inside the mask.  For the matrix of eq.~\eqref{MatrizMascara} to be of maximum rank (see
appendix~\ref{ApendiceMascara} for details), it is necessary that
\begin{equation}
\label{ecLimiteSimetriaMascara}
N_e \ge n_p/2, \,\,\, N_o \ge n_p/2,
\end{equation}
\noindent
due to the pixel symmetry, and
\begin{equation}
N_e+N_o \ge n_p+n_u,
\end{equation}
\noindent
so that there are at least as many spherical harmonics as pixels. The previous expressions allow us to find the minimum $\ell_{\max}$ that makes the matrix regular in a case with mask and $\SU$ symmetry.

We can also calculate the maximum rank of a matrix given the number of pixels, the mask, the number of pixels with and without a symmetric partner and the value of  $\ell_{\max}$. Transforming the matrix of eq.~\eqref{MatrizMascara} into diagonal blocks, in appendix \ref{ApendiceMascara} we show that the rank matrix is constrained by:
\begin{equation}
\label{RangoMascara}
\rank (\Y)   \le \min(n_p/2, N_e) + \min(n_p/2, N_o) + \min(n_u, L),
\end{equation}
were $L = N_e-\min(n_p/2, N_e) +  N_o-\min(n_o/2, N_e)$.}
The expression~(\ref{RangoMascara}),  $R_M$ hereafter, generalizes expression~(\ref{FINAL_Texto}) for any mask configuration.

\begin{figure} \begin{center}
\includegraphics[scale=0.25]{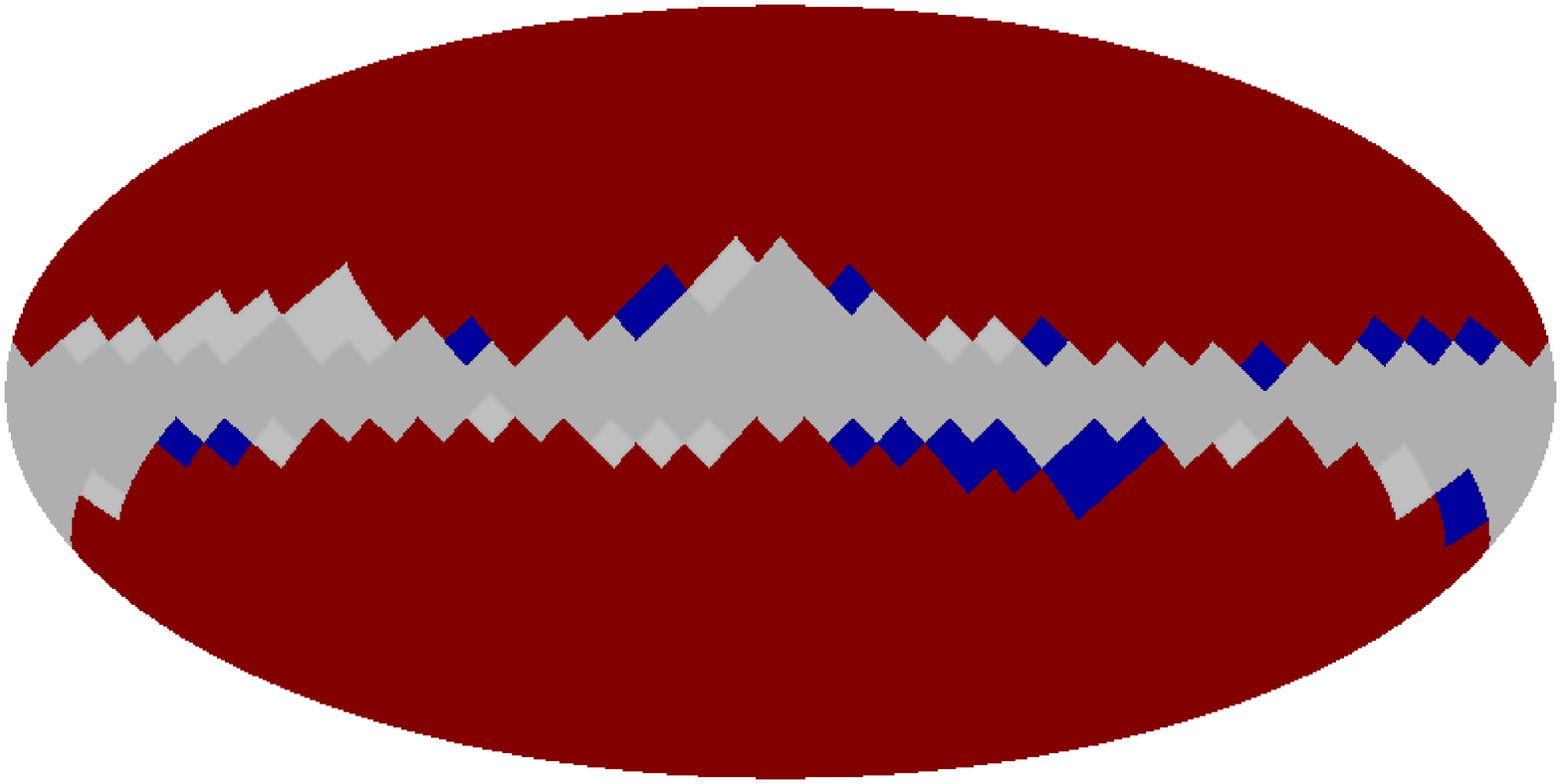}
\caption{Location of the pixels outside the \texttt{SEVEM} mask for $N_{\mathrm{side}}=8$. The pixels allowed by the mask with an $\SU$ symmetrical partner are red, while those without a symmetric partner are blue. The light and dark grey pixels are discarded by the mask. In particular, the light grey pixels show the points symmetrical to the unpaired blue ones.}
\label{FiguraMascara}
\end{center}
\end{figure}

As an example, let us consider the case of $N_{\mathrm{side}}=8$ in the \texttt{HEALPix} pixelization and the mask produced by the \texttt{SEVEM} component separation method \cite{2014A&A...571A..12P} corresponding to the first \emph{Planck} data release.\footnote{All Planck products are publicly available at the Planck Legacy Archive, \url{http://pla.esac.esa.int/pla}} The \texttt{SEVEM} mask removes 154 pixels of the sky at this resolution level. Among the 614 remaining pixels, 590 have a $\SU$ symmetrical pixel, whereas 24 do not have a symmetrical partner. Figure~\ref{FiguraMascara} shows the locations of paired and unpaired pixels for this case.

Table~\ref{TablaRangosMascara} shows the maximum expected rank $R_{\mathrm{th}}$ of the
covariance matrix according to $R_M$ and those calculated with
\textsc{Mathematica} using a realistic CMB power spectrum and beam ($R_N$). In addition, two
different precisions are considered: the native precision of the
machine and a 100-digit precision. We find that the rank estimated
with \textsc{Mathematica} coincides in all the considered cases with the
maximum allowed theoretical rank when using the higher
precision, whereas numerical errors affect the calculation using the
default precision.

\begin{table}
\begin{center}
\begin{tabular}{lccccccccccccc} \hline 
$l_{\max}$   & 21  & 22  & 23  & 24  & 25  & 26 \\ \hline
$R_{\mathrm{th}}$  & 480 & 525 & 572 & 614 & 614 & 614 \\
$R_N$  (default)  & 480 & 525 & 571 & 602 & 611 & 614 \\
$R_N$ (100)   & 480 & 525 & 572 & 614 & 614 & 614 \\ \hline	
\end{tabular}
\end{center}
\caption{Rank of $\C$ for the \texttt{HEALPix} pixelization at the
$N_{\mathrm{side}}=8$ resolution using the \texttt{SEVEM} mask for different
values of $\ell_{\max}$. $R_{\mathrm{th}}$ is the maximum theoretical rank according to $R_M$, and $R_N$ is
the rank of $\C$ calculated with
\textsc{Mathematica} using the default and 100-digit precisions.}
\label{TablaRangosMascara}
\end{table}

\section{Application to the QML method}\label{sec:QML}

As an example of application of the discussion above, let us consider
the Quadratic Maximum Likelihood (QML) method 
\cite{QML,1998PhRvD..57.2117B,teg01} introduced for the estimation of the
angular power spectrum. Different applications of the QML have been carried
out to obtain the CMB temperature and polarization power spectrum
(e.g. \cite{gru09,Planck2015_XI})
as well as the cross-correlation between the CMB and the Large Scale Structure
through the Integrated Sachs Wolfe effect \cite{sch12}.
The QML method gives unbiased, minimum
variance estimator of the CMB angular power spectrum, $\widehat{D}_{\ell}=
\ell(\ell+1) \widehat{C}_{\ell}/2\pi$, of a map $\x$ by means of a quadratic
estimator. The power can be computed through the following quantity
$y_{\ell}$:
\begin{equation}
\label{ecPotenciaEnEscalasAngulares}
y_\ell = \dfrac{1}{2} \dfrac{B_{\ell}^2}{\ell(\ell+1)/2\pi} \x^t\C^{-1}\mP^\ell \C^{-1}\x - n_{\ell},
\end{equation}
where $B_{\ell}$ takes into account the effect of the beam as well as
the pixel window function, $n_\ell$ is a correction term that removes
the noise bias and the matrix $\mP^{\ell}$ is defined in eq.~\eqref{CFormaHabitual}. The power in $y_{\ell}$ is a linear combination
of the \emph{true} power of the map:
\begin{equation}
\label{ecMezclaPotencia}
y_\ell= \sum_{\ell'} \F_{\ell \ell'} D_{\ell'}, 
\end{equation}
where $\F$ is the Fisher matrix computed in terms of the parameters
$D_{\ell}$. If $\F$ is regular, an estimator of $\widehat{D}_{\ell}$ can be defined as:
\begin{equation}
\label{DefinicionEstimador}
\widehat{D}_{\ell}= \sum_{\ell'} \F^{-1}_{\ell \ell'} y_{\ell'}.
\end{equation}
As seen from eq.~\eqref{ecPotenciaEnEscalasAngulares}, the estimator requires the inverse of the covariance matrix $\C$ and therefore the performance of the QML method depends critically on the regularity of that matrix.

As a working example, let us consider the case of \texttt{HEALPix}
pixelization at resolution $N_{\mathrm{side}}=32$ and full-sky. The total
number of pixels is 12288 and, according to the theoretical values given in
table~\ref{Tabla:RangosTeoricos}, it is necessary to sum up to
$\ell_{\max}=111$ in order to get the maximum rank of $\C$. However, from
table~\ref{RangosRuidoExperimental} we know that with this $\ell_{\max}$
the numerical matrix is not regular. A solution for this problem is to
add a \emph{small} amount of noise to make $\C$ regular. However, what does
\emph{small} mean in this context? If the artificial noise is too large,
the performance of the QML method will suffer degradation,
particularly at high multipoles, but if the noise level is too small
the matrix will pose numerical instabilities. Therefore the choice of
the correct noise level is not trivial. In this section we will study
in detail the effect of noise addition and investigate
the optimal \emph{small} amount of noise to be added.

In particular, we will study the value of the determinant $\lvert \C
\C^{-1} \rvert$, the statistical properties of the product $\eta =\x^t
\C^{-1} \x$ (which, since both the simulated CMB and noise are Gaussian, should follow a $\chi^2$ distribution with the same number of degrees of freedom as of pixels in the map), and the estimation and error bar of the power spectrum
given by the QML method after the addition of different levels of
noise. In order to do this, we first compute the matrix $\mS$ in
eq.~\eqref{MCRuido2} using the Planck best-fit model (and
including the beam and pixel window function) with $\ell_{\max} = 111$
and secondly we add isotropic white noise with variance $\sigma^2 =
10^{-f_n}$; CMB maps are then simulated with the \texttt{HEALPix}
package at $N_{\mathrm{side}}=32$, and noise is added following a normal
distribution with dispersion parameterised by $f_n$.

Table~\ref{TablaResumenQML} shows the value of the
determinant $\lvert \C \C^{-1} \rvert$ (fifth column) for different levels of
regularizing noise as well as the quantity $-\log \left( {\rm abs}
\left( 1-|\C \C^{-1}| \right) \right)$ (sixth column), introduced
in section~\ref{sec:Ruido}. It is apparent that for very low levels
of noise, the covariance matrix becomes singular and the determinant
produces values which are very far from unity. Conversely, for values of $f_n
\lesssim 5$, the determinant departs from its theoretical value
approximately in the sixth decimal place (or even better). Note that $f_n=5$
corresponds to a very small level of
noise compared to the signal (second column)
and, therefore, it is not expected to imply a degradation of the data. Of
course, the more noise we include, the closer the
determinant gets to unity, but at the price of degrading the data.

\begin{figure}
\begin{center}
\begin{tabular}{ccc}
\includegraphics{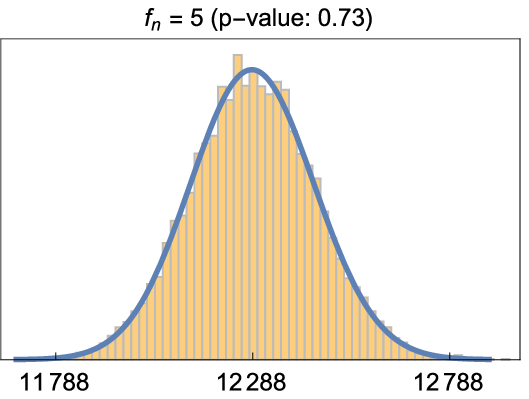} & \quad & 
\includegraphics{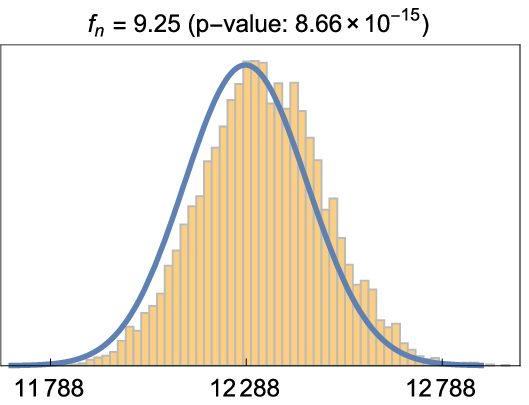} \\ 
\end{tabular}
\end{center}
\caption{Distribution of numerical values of $\x^t \C^{-1} \x$
compared with the theoretical distribution, for noise levels $f_n=$
$5$ and $9.25$ obtained from 10000 simulations.}
\label{GraficaTestChi2}
\end{figure}

The product $\eta =\x^t \C^{-1} \x$ should follow a $\chi^2$
distribution with 12288 degrees of freedom. However, if there are
numerical instabilities in $\C^{-1}$, this quantity will depart from
its theoretical distribution. In figure~\ref{GraficaTestChi2} we
show two histograms of $\eta$ obtained from 10000 simulations of CMB
plus noise with two different levels of noise in comparison with the
expected $\chi^2$ distribution. The $f_n=5$ noise results in a regular covariance matrix, and $f_n=9.25$ is in the limit in which the tests fail dramatically. It clearly shows how the distribution
of $\eta$ departs from the theoretical distribution when the level of
the noise is low. This can be quantified, for instance, by applying
the Cram\'er-von Mises goodness-of-fit hypothesis test
\cite{anderson1962}, which determines the $p$-value of the data to be consistent with the theoretical distribution. The $p$-values obtained for the different considered cases are given in table~\ref{TablaResumenQML} (fourth column).
As one would expect, the data are consistent with a $\chi^2$ distribution only when the covariance matrix is regularized by adding a certain
level of noise to the simulated data. 

\begin{figure}
\centering
\includegraphics{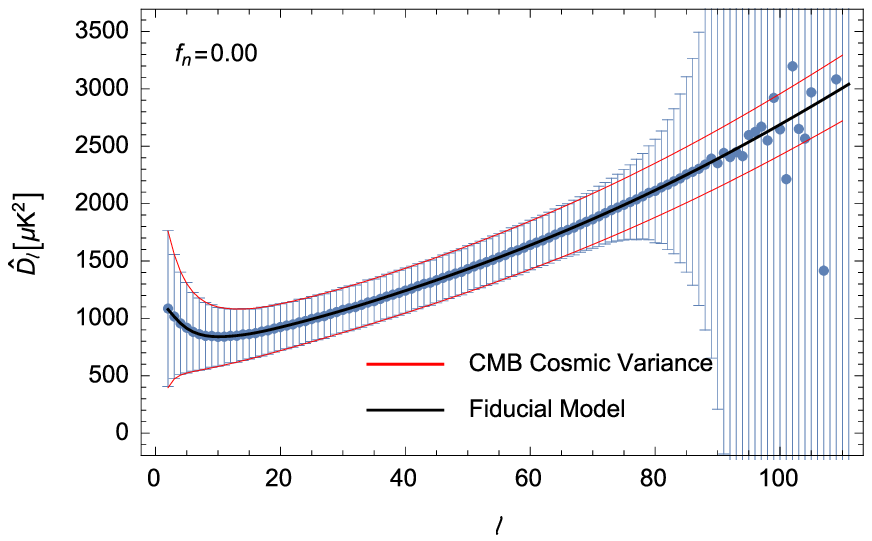} \vspace{2mm} \\
\includegraphics{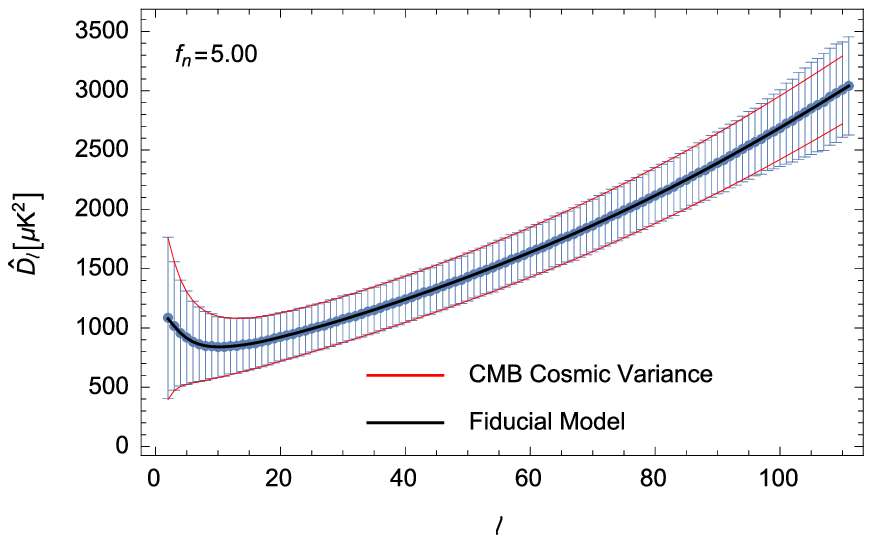} \vspace{2mm} \\
\includegraphics{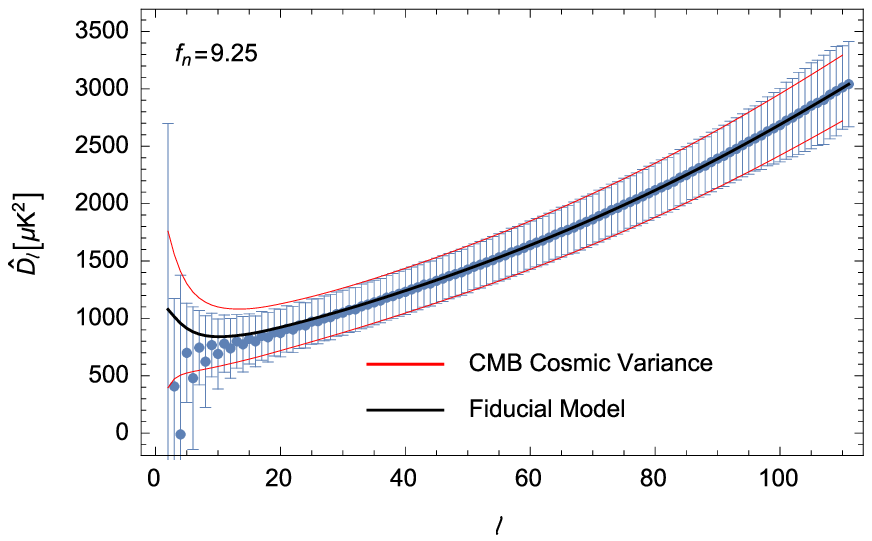} \\
\caption{Power spectrum estimation given by the QML method for three
different levels of noise, $f_n=\{0, 5, 9.25\}$.The average and
1$\sigma$ error bars have been obtained from a set of 10.000
simulations.}
\label{GraficaQML}
\end{figure}

So far, we have studied two generic statistics directly related to the
covariance matrix. We will now investigate how the possible singularity of
$\C$ propagates through the QML estimator. Figure~\ref{GraficaQML} shows the average
and $1\sigma$ error for $\widehat{D}_\ell$ when considering three different
levels of regularizing noise, namely, $f_n=$ 0, 5 and 9.25. If the
noise level is too small, the estimation of the low multipoles is systematically biased
downwards. If the noise is too high, the performance of the estimator
degrades at high $\ell$ and the error bars grow quickly. At intermediate
noise levels, the estimator is unbiased and the performance of the method is basically limited only
by cosmic variance, up to $\ell \approx 3
N_{\mathrm{side}}$.

To further quantify the performance of the QML method under the
presence of different levels of regularizing noise, we have calculated
the following quantity:
\begin{equation}
\label{ecDefinicionMedia} 
\left\langle \frac{\sigma_{\ell}^{QML}}{\sigma_{\ell}^{CV}} \right\rangle = 
\dfrac{1}{3 N_{\mathrm{side}}-1} \sum_{\ell=2}^{3 N_{\mathrm{side}}} \frac{\sigma_{\ell}^{QML}}{\sigma_{\ell}^{CV}},
\end{equation}
where $\sigma_{\ell}^{QML}$ corresponds to the dispersion of the QML
estimator and $\sigma_{\ell}^{CV}$ to that inferred from the cosmic
variance. This average ratio gives information on how
much the error in the estimation of the QML degrades with respect to
the optimal case. Table~\ref{TablaResumenQML} shows this ratio for different levels of
added noise. For a very low level of noise, the performance of the
estimator is clearly degraded due to the 
singularity of $\C$, while for a high level of noise, its effect in
increasing the error of the estimator at high multipoles also becomes
apparent. There is an intermediate region at which the error in the
estimation is very close to the error expected by cosmic variance,
i.e.,  $\langle \sigma_{\ell}^{QML}/\sigma_{\ell}^{CV}
\rangle \approx 1 $.

By looking simultaneously at the results for the
three previous tests, we can get a better idea of which
is the optimal level of noise to be added in order to regularize the
covariance matrix without degrading the data for the particular case
studied in this section. The hypothesis test based on $\x^t \C^{-1} \x$ and the determinant $|\C \C^{-1}|$ probe similar properties, since both of them focus directly on the inverse of $\C$.
As one would expect, the values in table~\ref{TablaResumenQML}
show that the results from both statistics are consistent. In particular, when the noise is
high, both tests show that the covariance matrix is regular, while,
when the noise is reduced, the determinant departs from unity and
the $p$-value of the Cram\'er-von Mises test gets close to
zero. Note that the more noise is added, the better the discriminant
is, but this does not indicate the level of noise at which the quality of the data starts to get compromised. Therefore, it is necessary to
look at other complementary statistics (such as the one based on the QML
method) in order to find the appropriate range of noise,
sufficiently high as to regularize the covariance matrix but also as
low as possible so that the data is not degraded.  By studying the
error bar in the estimation of the power spectrum with QML with respect to
that expected by cosmic variance, we find that very low values of
regularizing noise ($f_n=9$) can give an apparent good result in
this test but a bad value for the determinant. Thus, for
the sake of consistency and being conservative, one should choose a value of noise that achieves
both a good value for the determinant and a minimum error for the
QML method. In this sense, table~\ref{TablaResumenQML}  shows that values of $f_n$
between 3 and 5 are the most appropriate; the error bar criterion
gives the lowest value for the error size, the determinant departs
from unity at the sixth decimal position (in the worst case), and the
hypothesis test shows consistency with the expected theoretical
distribution.

\begin{table}
\begin{center}
\small{
\begin{tabular}{ccccrc} \hline
$f_n$ 	& $\%$ noise 		& $\langle \sigma_{\ell}^{QML}/\sigma_{\ell}^{CV} \rangle_{\ell}$ & $p$-value  & $  \lvert \C \C^{-1} \rvert $ & $-\log \left(  \text{abs}  \left( 1 - \lvert  \C \C^{-1} \rvert \right) \right)$ \\ \hline
0 	 	& 2.08 	& 2.54 	& 0.63 			& 1.00000000006 		& 10.22 \\
1 	 	& 0.66 	& 1.16 	& 0.93	 		& 1.0000000007 		& 9.13 \\
2 	 	& 0.21 	& 1.02 	& 0.66 	 		& 1.000000003 		& 8.49 \\
3 	 	& $6.6 \times 10^{-2}$	& 1.01 	& 0.73 	 		& 0.999999993 		& 8.16 \\
4 	 	& $2.1 \times  10^{-2}$ 	& 1.01 	& 0.73 	 		& 1.000001 			& 5.98 \\
5 	 	& $6.6 \times 10^{-3}$ 	& 1.01	& 0.73 	 		& 0.999998 			& 5.95 \\
6 	 	& $2.1 \times  10^{-3}$	& 1.01 	& 0.73 	 		& 1.000009 			& 5.05 \\
7 	 	& $6.6 \times 10^{-4}$	& 1.01 	& 0.60 	 		& 1.0002 				& 3.66 \\
8 	 	& $2.1 \times  10^{-4}$ 	& 1.01 	& 0.11 	 		& 0.97				& 1.50 \\
9 	 	& $6.6 \times 10^{-5}$	& 1.01 	& $\sim 10^{-15}$ & $1.27 \times  10^{36}$ 		& -36.10 \\
9.25 	& $4.9 \times  10^{-5}$  	& 1.13 	& $\sim 10^{-15}$ & $-1.12 \times  10^{299}$   	& -299.05 \\
10 	& $2.1 \times  10^{-5}$  	& $ \sim 7 \times  10^5 $ & $ \sim 10^{-14}$  & $\infty$  		& $-\infty$  \\ \hline
\end{tabular}}
\end{center}
\caption{Summary of statistical tests obtained from 10000 CMB
simulations (at $N_{\mathrm{side}}=32$) with different levels of
regularizing noise. The first column
shows the value of the noise parameter $f_n$, $\sigma_{noise}^{2}
= 10^{-f_n}$.  The second column gives the percentage of noise
with respect to the dispersion of the CMB signal, i.e., $100 \times
\sigma_{noise}/\sqrt{S_{ii}}$.  The third column shows the average
over $\ell$ of the errors on the estimation of $D_{\ell}$ divided by
those corresponding to the cosmic variance, 
eq.~\eqref{ecDefinicionMedia}. The $p$-value corresponds to
the probability that the quantity $\x^t \C^{-1}\x$ follows the
expected $\chi^2$ distribution according to the Cram\'er-von Mises
test. The two last columns show the results in relation to the
determinant of $\lvert \C \C^{-1} \rvert $.}
\label{TablaResumenQML}
\end{table}

For a generic application, one could give some general
considerations to proceed in order to have a regular covariance matrix
and to help establishing the appropriate level of regularizing noise,
if necessary, at each case:

\begin{enumerate}
	\item{Taking into account the number of pixels, the symmetries and if
	there is a mask, calculate the minimal theoretical
	multipole $\ell_{tm}$ up to which is necessary to sum in the
	calculation of $\C$ in order to get a regular matrix,
	using the adequate theoretical expression among the ones given in
	this paper.}
	\item{If the real value up to which we sum in the calculation of
	$\C$, $\ell_{\max}$, is lower than the previously determined multipole,
	$\ell_{tm}$, noise certainly needs to be added. Even if it is not
	lower, we might need to add some regularizing noise due to
	numerical errors.}
	\item{If necessary, add an amount of noise. Table~\ref{TablaResumenQML} may help to choose an initial guess.}
	\item{Calculate the determinant of $\C \C^{-1}$ and evaluate how much it deviates from unity. One should
	bear in mind that a very good determinant can also be the
	consequence of adding an excessive amount of noise.}
	\item{If the determinant is good, study the performance of the considered application (in our case
	the QML method) for this
	covariance matrix and compare it with the expected theoretical
	result.}
	\item{If either the value of the determinant or the performance of your
	application is not as good as expected, iterate by
	changing the amount of regularizing noise until both goals are achieved.}
\end{enumerate}

\section{Conclusions}
\label{sec:Conclusiones}

We have presented a comprehensive study of the regularity of the
covariance matrix $\C$ of a discretized field on the sphere. For a
general case, we have shown that a necessary condition for the
determinant of $\C$ to be different from zero is that the number of
considered harmonics is equal or greater than the number of
pixels. The presence of specific symmetries for the considered
pixelization or the use of a mask that excludes some regions of the
sphere also impose additional constraints on the rank of $\C$. Along
this work, five different expressions that establish limits on the
maximum rank achieved by the covariance matrix are presented; they
depend on the number of pixels, the number of spherical harmonics, the
kind of symmetries of the considered pixelization and the presence of
a mask.

When putting in practice these expressions for five different
pixelizations proposed within the context of CMB analysis (\texttt{Cube},
\texttt{Icosahedron}, \texttt{Igloo},  \texttt{GLESP} and \texttt{HEALPix}), we have found that the
particular properties of the location of the pixels can lead to a
reduction of the rank of the covariance matrix with respect to the
maximum rank allowed by the previously derived constraints.
Interestingly, among the five studied pixelizations, \texttt{HEALPix}
seems to present a lower reduction of the rank than the rest of
pixelizations, which should help to achieve the regularity of the
covariance matrix more easily.

We have also tested that numerical error propagation can give rise to
a loss of rank, which would produce a singular covariance matrix in
cases where this is not expected theoretically. By studying this
effect with different numerical precisions, it is possible to
differentiate whether the reduction of the rank is due to the
propagation of numerical errors or to an intrinsically singular covariance
matrix. As it is well known, a possible solution to mitigate the effect of
numerical errors is to add some level of uncorrelated noise in order to regularize
the covariance matrix. We have tested that even small levels of
noise can be sufficient, although the
particular value of the required noise will depend on the considered
case.

We have also considered a practical case in which the calculation of
the inverse of $\C$ is required, namely, the estimation of the CMB
temperature power spectrum using the QML estimator. In particular, we
have considered simulations containing CMB and different levels of
noise using the \texttt{HEALPix} pixelization at resolution $N_{\mathrm{side}}
= 32$. The optimal level of noise which needs to be included to obtain
a regular matrix without degrading the quality of the data has been
investigated in detail. We have studied particularly the behaviour
of three different statistics in order to evaluate the quality of the results
versus different levels of regularizing noise: the value of the
determinant of $\C\C^{-1}$, a hypothesis test to study the distribution
of the quantity $\eta =\x^t \C^{-1} \x$ (which should follow a
$\chi^2$ distribution if $\x$ represents Gaussian random fields) and the increment in the error of the
estimation of the power spectrum with respect to the ideal case (given
by the cosmic variance). The first conclusion from this particular
exercise is that some level of noise, even if small, must be
introduced in order to get a regular matrix, since numerical errors
make the matrix singular even if this was not expected from the
theoretical point of view. As the level of noise increases, the
determinant and the hypothesis test indicate that the covariance
matrix becomes more regular (for instance, it is observed that the
determinant of $\C\C^{-1}$ departs less from unity). However,
increasing the level of noise also degrades the quality of the data,
which is reflected in the increment of the error in the estimation of
the CMB power spectrum with the QML estimator. Also note that the
error of this estimator can give good results even if the determinant
criterion is bad. Therefore, for the sake of consistency, it becomes
necessary to look simultaneously at all the statistics in order to
establish the optimal level of noise for the error in the QML estimation to be as small as possible, 
while ensuring the covariance matrix is well behaved. For this particular case, this
can be achieved with a level of noise as small as $\sim$ 0.01 per cent
with respect to the CMB dispersion (see table~\ref{TablaResumenQML}),
although the particular value will depend on the considered case as
well as on the required precision for the covariance matrix.

Finally, some general guidelines concerning the steps to follow to get a regular matrix for a 
generic application are given in section~\ref{sec:QML}.



\appendix
\section{\texorpdfstring{Study of the determinant of \boldmath $\C$}{Study of the determinant of C}}
\label{Determinante}

According to eq.~\eqref{FinalDesarrollo}, we can expand the determinant of $\C$ as a sum of determinants of matrices whose columns are the spherical harmonics calculated at the different pixels of the image. If one determinant in the sum is not null, there will be other $n! -1$ terms that are not zero, corresponding to the permutations of its columns. However, in principle, it could happen that the sum of all these elements is zero. In this appendix we will calculate this sum and show that the result is a real, positive number.

Let us suppose that we have a particular collection of values $\left\lbrace \mu_1, \mu_2, \ldots, \mu_n \right\rbrace$ such that the corresponding determinant in eq.~\eqref{FinalDesarrollo} is not null; of course, the indices must be different and correspond to a collection of linearly independent spherical harmonics. For the sake of simplicity, let us reassign labels to the spherical harmonics in such a way that now $Y_{\mu_i}$ is labelled as $Y_{i}$, so we have a matrix whose $n$ columns are the $n$ spherical harmonics $Y_{i}, i=1, \ldots n$ evaluated in the $n$ pixels of the map. Let us define D and $\M$:
\begin{equation}
\label{DefinicionD}
D \equiv  \det(\M) \equiv
\det \left(  
\begin{array}{cccc}
Y_{11}  & Y_{21} & \cdots & Y_{n1} \\
Y_{12}  & Y_{22} & \cdots & Y_{n2} \\
\vdots & \vdots & & \vdots \\
Y_{1n}  & Y_{2n} & \cdots & Y_{nn} \\
\end{array} \right).
\end{equation}
In the concatenated summations in eq.~\eqref{FinalDesarrollo} there are another $n!-1$ no null terms given by permutations of the columns of the matrix in eq.~\eqref{DefinicionD}, whose determinants take values $\pm D$ depending on the signature of the permutation. Let $S$ be the contribution to the determinant of $\C$ of the terms coming from those permutations, let $P_n$ be the collection of permutations of $n$ elements, and let us also relabel $C_{\mu_i}$ in eq.~\eqref{FinalDesarrollo} as $C_{i}$. The summation extends over the permutations $P_n$, and:
\begin{align}
\label{SumaPermutaciones}
S & = \sum_{\sigma \in  P_n} \left[ \prod_{i=1}^n C_i \right]  \nonumber
\left[ \prod_{i=1}^n Y_{\sigma_i i}^* \right]  \mathrm{sign} \left(
\sigma \right) D \\
& = D \left[ \prod_{i=1}^n C_i \right] \det \left( {\M}^* \right) \nonumber \\
& = \left[ \prod_{i=1}^n C_i \right] |D|^2.
\end{align} 
Note that in order to get to the second line of the equation, we have taken out
of the sum the constant $D$ as well as the factors coming from the power spectrum
that multiplies each permutation in eq.~\eqref{FinalDesarrollo}, since
these factors only depend on the spherical harmonics in the
permutation and not on its order. In this way we have identified the
remaining sum as Leibniz's formula for the determinant of
$\M^*$. Taking into account that determinant and conjugate commute, we
get to the final expression of the previous equation.

Therefore, assuming that there are enough non-zero $C_{\ell}$'s, if
we are able to find a collection of spherical harmonics such that $D\ne0$,
we have:
\begin{equation}
\det(\C) > 0.
\end{equation}

Finally, let us calculate the full value of the determinant of
$\C$. If we consider $n$ points on the sphere and $N$ spherical
harmonics, we can find as many different sets of $n$ spherical harmonics as the number of $n$-combinations of an
$N$-set. Since the expression~\eqref{SumaPermutaciones} gives the generic contribution to the determinant of $\C$ of a set of $n$ harmonics, 
if we call $\sigma$ to one of the combinations, and $S_{\sigma}$ to
the sum in eq.~\eqref{SumaPermutaciones} for this combination, then:
\begin{equation}
\label{FinalDetC}
\det(\C) = \sum_{\sigma} S_{\sigma} = \sum_{\sigma} \left( \prod_{i=1}^{n} C_{\sigma_i} \right) |D_{\sigma}|^2.
\end{equation}

\section{\texorpdfstring{\boldmath $\SU$ symmetry: $\vvec{r} \rightarrow -\vvec{r}$}{SI symmetry}}
\label{Simetria1}

The spherical harmonics satisfy
\begin{equation}
\label{SimetriaY}
Y_{\ell m}(-z, \phi + \pi) =  {(-1)}^{\ell}Y_{\ell m}(z, \phi).
\end{equation}
If for each pixel with coordinates $(z, \phi)$  there exists a pixel with coordinates $(-z, \phi + \pi) $, the matrix  
in eq.~\eqref{MatrizGenericaB} simplifies notably. This can be easily seen
considering the simple case with $\ell=1$ and four pixels:
\begin{equation}
\label{SimetriaA}
\Y = 	\left(  
\begin{array}{ccc}
Y_{1}^{-1}(\vvec{r}_1)  & Y_{1}^{0}(\vvec{r}_1)  & Y_{1}^{1}(\vvec{r}_1) \\
Y_{1}^{-1}(\vvec{r}_2)  & Y_{1}^{0}(\vvec{r}_2)  & Y_{1}^{1}(\vvec{r}_2) \\
Y_{1}^{-1}(\vvec{r}_3)  & Y_{1}^{0}(\vvec{r}_3)  & Y_{1}^{1}(\vvec{r}_3) \\
Y_{1}^{-1}(\vvec{r}_4)  & Y_{1}^{0}(\vvec{r}_4)  & Y_{1}^{1}(\vvec{r}_4) \\
\end{array} \right).
\end{equation}
If the pixels are opposed pairwise, 
$\vvec{r}_3 = - \vvec{r}_1$ y $\vvec{r}_4 = - \vvec{r}_2$, then the matrix of eq.~\eqref{SimetriaA} is a rank 2 matrix, since we have only two linearly
independent rows:
\begin{equation}
\label{SimetriaA2}
\Y = 	\left(  
\begin{array}{rrr}
Y_{1}^{-1}(\vvec{r}_1)  & Y_{1}^{0}(\vvec{r}_1)  & Y_{1}^{1}(\vvec{r}_1) \\
Y_{1}^{-1}(\vvec{r}_2)  & Y_{1}^{0}(\vvec{r}_2)  & Y_{1}^{1}(\vvec{r}_2) \\
-Y_{1}^{-1}(\vvec{r}_1)  & -Y_{1}^{0}(\vvec{r}_1)  & -Y_{1}^{1}(\vvec{r}_1) \\
-Y_{1}^{-1}(\vvec{r}_2)  & -Y_{1}^{0}(\vvec{r}_2)  & -Y_{1}^{1}(\vvec{r}_2) \\
\end{array} \right).
\end{equation}

With the help of this property we can notably
simplify the matrix $\Y$. But let us first define an operator that
will simplify the notation in the coming expression. Suppose that we have
selected a certain pixel collection, indexed from $1$ to $k$, and a set
of harmonics, represented by $Y_{LM}$, that contains $U$ elements
indexed from $1$ to $U$. The operator $\otimes$ constructs the block
obtained by applying all the harmonics of the set on the selected
pixels:
\begin{equation}
Y_{LM} \otimes
\left( 
\begin{array}{c}
\vvec{r}_1 \\  \vdots \\ \vvec{r}_k \\
\end{array}
\right) _{k \times U} 
\equiv
\left(\begin{array}{ccccc}
Y_{\ell_1 m_1}(\vvec{r}_{1})  &  \cdots  &  Y_{\ell_U m_U}(\vvec{r}_{1})    \\ 
\vdots & & \vdots \\ 
Y_{\ell_1 m_1}(\vvec{r}_{k})  &  \cdots  &  Y_{\ell_U m_U}(\vvec{r}_{k})    \\    
\end{array}
\right).
\end{equation}
Going back to $\Y$ and considering the case of $n$ pixels
in the pixelization that fully accomplish the $\SU$ symmetry,
$\vvec{r}_{i + n/2} = - \vvec{r}_i, \; i=1\ldots n/2$, if we reorder the
spherical harmonics in such a way that the ones with even $\ell$ occupy
the first columns and those with odd $\ell$ occupy the last columns, $\Y$
becomes:
\begin{equation}
\Y = \left(
\begin{array}{cc}
Y_{\ell_{e}} \otimes \left(\begin{array}{c}
\vvec{r}_1 \\  \vdots \\ \vvec{r}_{n/2}
\end{array}
\right)_{n/2 \times N_e} 
&  
Y_{\ell_{o}} \otimes \left(\begin{array}{c}
\vvec{r}_1 \\  \vdots \\ \vvec{r}_{n/2}
\end{array}
\right)_{n/2 \times N_o} 
\\

\\
Y_{\ell_{e}} \otimes \left(\begin{array}{c}
\vvec{r}_{n/2+1} \\  \vdots \\ \vvec{r}_{n}
\end{array}
\right)_{n/2 \times N_e} 
& 
Y_{\ell_{o}} \otimes \left(\begin{array}{c}
\vvec{r}_{n/2+1} \\  \vdots \\ \vvec{r}_{n}
\end{array}
\right)_{n/2 \times N_o} 

\end{array}
\right),
\end{equation}
\noindent
where $\ell_e$ and $\ell_o$ represent the set of spherical harmonics
with even and odd $\ell$ respectively. The corresponding numbers of harmonics in each block are given by $N_e$ and
$N_o$ (see eq.~\eqref{eq:eses}).

Taking into account the symmetry of pixels and spherical harmonics, we have:
\begin{equation}
\Y =  \left( 
\begin{array}{cc}
Y_{\ell_{e}} \otimes \left(\begin{array}{c}
\vvec{r}_1 \\  \vdots \\ \vvec{r}_{n/2}
\end{array}
\right)_{n/2 \times N_e} 
& 
Y_{\ell_{o}} \otimes \left(\begin{array}{c}
\vvec{r}_1 \\  \vdots \\ \vvec{r}_{n/2}
\end{array}
\right)_{n/2 \times N_o} 
\\
\\
Y_{\ell_{e}} \otimes \left(\begin{array}{c}
\vvec{r}_{1} \\  \vdots \\ \vvec{r}_{n/2}
\end{array}
\right)_{n/2 \times N_e} 
& 
-Y_{\ell_{o}} \otimes \left(\begin{array}{c}
\vvec{r}_{1} \\  \vdots \\ \vvec{r}_{n/2}
\end{array}
\right)_{n/2 \times N_o} 

\end{array}
\right).
\end{equation}

It is possible to substitute the first $n/2$ rows $r_i$ by the linear
combination $\frac{1}{2} (r_i +r_{i+n/2}), \; i=1,\ldots,
n/2$. Similarly, the
second half of rows can be replaced by $\frac{1}{2}
(r_{i-n/2} -r_{i}), \; i=n/2+1, \ldots, n$. With this transformation,
one finds:
\begin{equation}
\label{Matriz2Bloques}
\Y \rightarrow \left( 
\begin{array}{cc}
Y_{\ell_{e}} \otimes \left(\begin{array}{c}
\vvec{r}_1 \\  \vdots \\ \vvec{r}_{n/2}
\end{array}
\right)_{n/2 \times N_e} 
& 
\0_{n/2 \times N_o} 
\\
\\
\0_{n/2 \times N_e} 
& 
Y_{\ell_{o}} \otimes \left(\begin{array}{c}
\vvec{r}_{1} \\  \vdots \\ \vvec{r}_{n/2}
\end{array}
\right)_{n/2 \times N_o} 

\end{array}
\right).
\end{equation}

Therefore, the rank of the matrix is equal to the sum of the ranks of
the two diagonal blocks above. Each of these ranks is less or equal
than the minimum between the number of rows ($\leq n/2$) and the
number of columns ($N_e$ or $N_o$), i.e., 
\begin{equation}\label{FINAL}
\rank(\Y) \le \min \left(N_e, n/2 \right) + \min \left(N_o, n/2 \right),
\end{equation}
which corresponds to the constraint $R_1$ presented in section~\ref{sec:s1}.

\section{\texorpdfstring{\boldmath $\SD$ symmetry: $\phi \rightarrow \phi+\pi$}{SII symmetry}}
\label{Simetria2}

Another interesting property of the spherical harmonics is:
\begin{equation}
\label{AESimetriaM}
Y_{\ell m}(z, \phi+\pi)  = {(-1)}^m Y_{\ell m}(z, \phi).
\end{equation}
Let us assume that we have a pixelization that satisfy the previous
symmetry $\SU$. 
If for each point in the pixelization with coordinates
$(z,\phi)$ there is another point with coordinates $(z,\phi+\pi)$, then it is possible to further simplify the
matrix of eq.~\eqref{Matriz2Bloques}. In order to do this, one should first
note that the non-zero blocks are
evaluated only on the first half of the pixels. 
All pixels with $z \ne 0$ that are included in these blocks have
their symmetric pair $\phi \rightarrow \phi+\pi$ also in the first half
of pixels, which allows one to use symmetry $\SD$ to reduce the size of
the non-zero blocks. However, the case with $z=0$ needs to be addressed separately. For these points,
the corresponding $\phi \rightarrow \phi+\pi$ pixel satisfies also $\vvec{r}
\rightarrow -\vvec{r}$, and therefore it has been
removed out of the matrix during the transformations that led to
the matrix of eq.~\eqref{Matriz2Bloques}. Therefore, for those positions with $z=0$, the pixel symmetric
with respect to the $\SD$ symmetry is not present in the non-zero blocks and can not
be used to further simplify the matrix of eq.~\eqref{Matriz2Bloques}.

Fortunately, there is another interesting property of the spherical
harmonics that can be used in this case. If $\ell$ is even and $m$ is odd, or if
$\ell$ is odd and $m$ is even, then $Y_{\ell m}(z=0,\phi)=0$.
Let us assume that when applying the symmetry $\SU$, we
have kept the first $n/2$ pixels (those with $z>0$, and those with $z=0$
and $ 0 \le \phi < \pi$). Let us now reorder the spherical
harmonics (i.e., the columns of matrix $\Y$) in a sequence such that the values $(\ell, m)$ are arranged in
the following combinations: (even, odd), (even, even), (odd, odd), and
finally (odd, even). After this rearrangement, the even $\ell$ block of
matrix of eq.~\eqref{Matriz2Bloques} becomes:
\begin{equation}
\label{BloqueLPar}
\Y_{e} = \left( 
\begin{array}{cc}
Y_{\ell_{e} m_o} \otimes \left(\begin{array}{c}
\vvec{r}_1 \\  \vdots \\ \vvec{r}_{k}
\end{array}
\right)_{k \times N_{eo}} 
& 
Y_{\ell_{e} m_e} \otimes \left(\begin{array}{c}
\vvec{r}_1 \\  \vdots \\ \vvec{r}_{k}
\end{array}
\right)_{k \times N_{ee}}
\\ 
\\
\0_{u \times N_{eo}} 
& 
Y_{\ell_{e} m_e} \otimes \left(\begin{array}{c}
\vvec{r}_{k+1} \\  \vdots \\ \vvec{r}_{k+u}
\end{array}
\right)_{u \times N_{ee}} 
\end{array}
\right),
\end{equation}
\noindent
where we have also separated the rows corresponding to $k$ pixels with
$z>0$ from the $u$ positions with $z=0$, thus $k+u=n/2$.

$N_{eo}$ is the number of spherical harmonics with even $\ell$ and odd
$m$, and $N_{ee}$ is the number of spherical harmonics with even $\ell$
and even $m$. Using eq.~\eqref{AESimetriaM} and dividing the $k$ pixels
with $z>0$ in two halves, one with $0 \le \phi < \pi$ and another one with
$\pi \le \phi < 2 \pi$, the block of eq.~\eqref{BloqueLPar} can be
transformed:
\begin{equation}
\label{BloqueLPar2}
\Y_e \rightarrow	\left( 
\begin{array}{cc}
Y_{\ell_{e} m_o} \otimes \left(\begin{array}{c}
\vvec{r}_1 \\  \vdots \\ \vvec{r}_{k/2}
\end{array}
\right)_{k/2 \times N_{eo}} 
& 
\0_{k/2 \times N_{ee}}
\\ 
\\
\0_{k/2 \times N_{eo}} 
& 
Y_{\ell_{e} m_e} \otimes \left(\begin{array}{c}
\vvec{r}_1 \\  \vdots \\ \vvec{r}_{k/2}
\end{array}
\right)_{k/2 \times N_{ee}}
\\ 
\\
\0_{u \times N_{eo}} 
& 
Y_{\ell_{e} m_e} \otimes \left(\begin{array}{c}
\vvec{r}_{k+1} \\  \vdots \\ \vvec{r}_{k+u}
\end{array}
\right)_{u \times N_{ee}} 
\end{array}
\right),
\end{equation}
where we have operated in an analogous way to that used in the transformation that
led to the matrix of eq.~\eqref{Matriz2Bloques}. The odd $\ell$ block of this matrix
can also be simplified in the same manner, until $\Y$ is finally reduced
to a four diagonal block matrix. The total rank is the sum of the
ranks of the four blocks:
\begin{align}
\nonumber
\rank(\Y) & \le  \min (N_{eo}, k/2) + \min (N_{ee}, k/2+u) \\
&  + \min (N_{oo}, k/2+u) + \min (N_{oe}, k/2),
\end{align}
\noindent
where  $N_{oo}$ y $N_{oe}$ are the number of spherical harmonics with odd $\ell$ and $m$ and with odd $\ell$ and even $m$, respectively.

\section{\texorpdfstring{\boldmath $\ST$ symmetry: $\phi \rightarrow \phi+\pi/2$}{SIII symmetry}}
\label{Simetria3}

A third interesting symmetry relation of the spherical harmonics is:
\begin{equation}
\label{RelacionM2}
Y_{\ell m}\left(\theta, \phi+\pi/2\right) =   i^m Y_{\ell m}\left(\theta, \phi\right). 
\end{equation}
Again, let us assume that we have a pixelization that satisfy the
symmetries $\SU$ and $\SD$ and, therefore, we can transform the matrix
$\Y$ into a matrix with four diagonal blocks as the one described in eq.~\eqref{BloqueLPar2}.  If, in addition, for each point in the
pixelization with coordinates $(z,\phi)$ there is another point with
coordinates $(z,\phi+\pi/2)$, it is then possible to subdivide each block into two diagonal sub-blocks. To do that, we need to rearrange the pixels and spherical harmonics properly, and take linear combinations similar to the ones described in appendix~\ref{Simetria2}. We show as an example the transformation of the block of the spherical harmonics of even $\ell$ and odd $m$ of eq.~\eqref{BloqueLPar2}:
\begin{equation}
\label{BloqueLPar3}
\Y_{eo} = Y_{\ell_{e} m_{o}} \otimes \left(  \begin{array}{c}
\vvec{r}_1 \\  \vdots \\ \vvec{r}_{k/2}
\end{array}
\right)_{k/2 \times N_{eo}},
\end{equation}  
\noindent
into:
\begin{equation}  
\Y_{eo} \rightarrow \left(
\begin{array}{cc}
Y_{\ell_{e} m_{1}} \otimes \left(\begin{array}{c}
\vvec{r}_1 \\  \vdots \\ \vvec{r}_{k/4}
\end{array}
\right)_{k/4 \times N_{e1}} 
& 
\0_{k/4 \times N_{e3}} \\
\0_{k/4 \times N_{e1}} 
&  Y_{\ell_{e} m_{3}} \otimes \left(\begin{array}{c}
\vvec{r}_1 \\  \vdots \\ \vvec{r}_{k/4}
\end{array}
\right)_{k/4 \times N_{e3}}
\end{array} 
\right).  
\end{equation}
\noindent
$N_{e1}$ and $N_{e3}$ are the number of spherical harmonics with
$i^m=i$ and $i^m=-i$, respectively, and $m_{1}$ and $m_{3}$ indicate
the corresponding sets of indices. Note that, in this case, to obtain
zero blocks the imaginary unit $i$ has to be introduced in the linear
combinations of rows.  The block corresponding to the spherical
harmonics with even $\ell$ and $m$ in eq.~\eqref{BloqueLPar2}, and those
with odd $\ell$ can be similarly transformed. Finally, the matrix is
divided into eight diagonal blocks whose total rank is constrained by
eq.~\eqref{Formula4A}.

\section{Rank expression under the presence of a mask}\label{ApendiceMascara}

In order to prove the constraint given by eq.~\eqref{RangoMascara}, we start
from the full matrix $\Y$:
\begin{equation}
\label{MatrizYEnApendice}
\Y = 	\left(  
\begin{array}{ccc}
Y_{1,1} & \cdots  & Y_{N,1} \\
\vdots  &         & \vdots  \\
Y_{1,n} & \cdots  & Y_{N,n} \\
\end{array} 
\right).
\end{equation}
Let us consider that we have $n$ valid pixels allowed by the mask
that correspond to $n_p$ paired pixels (i.e.\ that fulfil the $\SU$ symmetry)
and $n_u$ unpaired pixels, $n = n_p + n_u$. Let us also assume that
there are $N_e$ and $N_o$ spherical harmonics
of even and odd $\ell$ respectively. By properly ordering the
spherical harmonics and the pixels, and after taking linear combinations of the $n_p$
rows as we did in appendix~\ref{Simetria1}, the matrix of eq.~\eqref{MatrizYEnApendice}
can be converted into:
\begin{equation}
\Y \rightarrow \left( 
\begin{array}{cc}
{\Y}_{n_p/2 \times N_e}^{even} 
& 
\0_{n_p/2 \times N_o} 
\\ \\
\0_{n_p/2 \times N_e} & {\Y}_{n_p/2 \times N_o}^{odd} 
\\ \\
{\Y}_{n_u \times N_e}^{even}  &  {\Y}_{n_u \times N_o}^{odd} 
\end{array}
\right)
\equiv
\left(\begin{array}{cc}
{\Y}_{p}^{e} 
& 
\0 
\\ \\
\0 & {\Y}_{p}^{o} 
\\ \\
{\Y}_{u}^{e}  &  {\Y}_{u}^{o} 
\end{array}
\right).
\end{equation}
If the rank of the block ${\Y}_{p}^{e}$ is $R_{e}$ and the one
of ${\Y}_{p}^{o}$ is $R_{o}$, we have:
\begin{equation}
\label{LimiteReRo}
R_{e} \le \min(n_p/2, N_e), \qquad R_{o} \le \min(n_p/2, N_o),
\end{equation}
since the rank of each block has to be less or equal than the minimum
between the number of rows and the number of columns.

Taking linear combinations, the blocks ${\Y}_{p}^{e}$ and
${\Y}_{p}^{o}$ of the previous matrix can be transformed into diagonal blocks with as
many no null elements as their respective ranks, $R_{e}$ and $R_{o}$.
Showing it explicitly for ${\Y}_{p}^{e}$, and
substituting ${\Y}_{p}^{o}$ by its diagonal form
$\D_{p}^{o}$, we have:
\begin{equation}
\Y \rightarrow	\left( 
\begin{array}{cc}
\begin{array}{cc}
\begin{array}{cccc}
d^{e}_{11} 	& 0 	& \cdots	& 0 \\
0		& d^{e}_{22} 	& \cdots	& 0 \\
\vdots	& \vdots		& \ddots	& \vdots \\
0		& 0				& \cdots	& d^{e}_{R_{e}R_{e}} \\
\end{array}
&	\0 \\ 
\\
\0 				& 	\0 \\
\end{array}
& 
\0 
\\ \\
\0 & 
\begin{array}{cc}
\D_{p}^{o}		&	\0 \\
\0				&	\0 \\
\end{array}
\\ \\
{\Y}_{u}^{e}  &  {\Y}_{u}^{o} 
\end{array}
\right).
\end{equation}
Note that some of the zero blocks around
${\D}_{p}^{e}$ and ${\D}_{p}^{o}$ will not appear if
the equality is fulfilled in eq.~\eqref{LimiteReRo}.

Furthermore, using the diagonal blocks, we can obtain $R_e$ zeros in
the ${\Y}_{u}^{e} $ block and $R_o$ zeros in the
${\Y}_{u}^{o}$ block. Therefore, the matrix can be further
simplified:
\begin{equation}
\Y \rightarrow	\left( 
\begin{array}{cc}
\begin{array}{cc}
\D_{p}^{e} 	&	\0 \hspace{3mm} \\
\0 			& 	\0 \hspace{3mm} \\
\end{array}
& 
\0 
\\ \\
\0 & 
\begin{array}{cc}
\D_{p}^{o}		&	\0  \hspace{3mm} \\
\0				&	\0  \hspace{3mm} \\
\end{array}
\\ \\
\begin{array}{cc}
\hspace{0.5mm} \0 & {\Y'}_{u}^{e} \\
\end{array}
&  
\begin{array}{cc}
\hspace{0.5mm} \0 & {\Y'}_{u}^{o} \\
\end{array}

\end{array}
\right),
\end{equation}
where the ${\Y'}_{u}^{e}$ and ${\Y'}_{u}^{o}$ blocks have
$N_e-R_e$ columns and $N_o-R_o$ columns, respectively.
Finally, reordering the rows and columns, we obtain:
\begin{equation}
\label{ecMatrizFinalBloquesMAscara}
\Y \rightarrow	\left(
\begin{array}{cccc}
\D_{p}^{e}	& \0 & \0 & \0 \\
\0	& \D_{p}^{o} & \0 & \0 \\
\0	&  \0 & {\Y'}_{u}^{e} & {\Y'}_{u}^{o} \\
\0 & \0 & \0 & \0  \\
\end{array} \right).
\end{equation}
Inspecting this matrix, we can easily derive the constraint given in eq.~\eqref{ecLimiteSimetriaMascara}. For the $\rank(\C)$ to be maximum,
i.e. $\rank(\C)=\rank(\Y)=n$, $\Y$ needs to have $n$ independent
rows. Therefore, in this case, we would get a matrix like the one of eq.~\eqref{ecMatrizFinalBloquesMAscara} but without the last row of zero
blocks. Moreover, given that the block $\D_{p}^{e}$ has $n_p/2$
rows, it must also have at least $n_p/2$ linear independent columns in
order to achieve the maximum rank. Since all of its columns are constructed in terms
of spherical harmonics of even $\ell$, then $N_{e} \ge n_p/2$
must be satisfied. Similarly, the same reasoning applies to block
$\D_{p}^{o}$, finding $N_{o} \ge n_p/2$.

Let us now study the rank of the matrix in eq.~\eqref{ecMatrizFinalBloquesMAscara} for any $\ell_{\max}$. Its rank is the sum of the ranks of the blocks $\D_{p}^{e}$, $\D_{p}^{o}$ and $({\Y'}_{u}^{e} \, {\Y'}_{u}^{o})$. Taking into account eq.~\eqref{LimiteReRo} and
since the block $({\Y'}_{u}^{e} \, {\Y'}_{u}^{o})$ has
$n_u$ rows and $L^*=N_e-R_e+N_o-R_o$ columns, we get the limit:
\begin{align}
\rank (\Y)  & \le R_{e} + R_{o} + \min(n_u, L^*) \nonumber \\
& \le  \min(n_p/2, N_e) + \min(n_p/2, N_o) + \min(n_u, L^*).
\label{RangoMascaraLEstrella}
\end{align}
However this expression depends through $L^*$ on the ranks $R_e$ and $R_o$ which
are, in principle, unknown. It would be more convenient to have a
more general constraint which did not depend on these ranks. With this
aim, let us consider a new parameter, $L$, defined as:
\begin{equation}
L = N_e-\min(n_p/2, N_e) +  N_o-\min(n_p/2, N_o).
\end{equation}
Taking into account eq.~\eqref{LimiteReRo}, it becomes apparent
that $L^*\ge L$ and $\min(n_u, L^*) \ge \min(n_u, L)$. Therefore, it
is not obvious whether the expression~(\ref{RangoMascaraLEstrella}) is still valid
when replacing $L^*$ by $L$ (except in the trivial case $L^*=L$).

Let us study in more detail the case $L^* > L$. We suppose $L^* = L +
t$, with $t \ge 1$. In this case, $R_e$ or $R_o$ (or both) are lower
than their possible maximum value given by the geometrical dimensions
of the corresponding blocks. For the sake of simplicity, and without
loss of generality, let us assume that $R_e = \min(n_p/2, N_e) -
t$ (note that if both $R_e$ and $R_o$ are lower than their maximum
geometrical value, the same reasoning would be done twice, with $t_e$ and $t_o$, $t=t_e+t_o$). In this situation, we have:
\begin{align}
\label{CasoReBajoT}  
\rank ( \Y ) & \le R_e + R_o + \min(n_u, L^*) \nonumber \\
& \le \min(n_p/2, N_e) - t + \min(n_p/2, N_o)  \nonumber \\
&  + \min(n_u, L^*). 
\end{align}
Let us now focus on the value of the last term, $\min(n_u, L^*)$, and
consider the two possible complementary cases $n_u \ge L^*$ and $n_u < L^*$. 

\begin{enumerate}

	\item{
	If $n_u \ge L^*$, then $\min(n_u, L^*) = L^* = L + t$, and $\min(n_u, L)=L$. Expression~(\ref{CasoReBajoT}) becomes:}
\begin{align}
\label{CasoReBajoTAFinal}  
\rank( \Y ) & \le \min(n_p/2, N_e) - t + \min(n_p/2, N_o) + L + t \nonumber \\
& =  \min(n_p/2, N_e) + \min(n_p/2, N_o) + \min(n_u, L).
\end{align}
\item{Conversely, if $n_u < L^*$, expression~(\ref{CasoReBajoT}) becomes:}
\begin{align}
\label{CasoReBajoTB}  
\rank ( \Y ) & \le  \min(n_p/2, N_e) + \min(n_p/2, N_o) + n_u - t. 
\end{align}
Since $L < L^*$, if $n_u < L^*$ we have again two complementary possibilities: $n_u
\le L$ and $n_u > L$.

\begin{enumerate}
	\item{
	Let us first suppose that $n_u \le L$, which
	implies $\min(n_u, L) = n_u$. In this case, from eq.~\eqref{CasoReBajoTB} we
	have:}
\begin{align}
\label{CasoReBajoTBC1}  
\rank (\Y) & < \min(n_p/2, N_e) + \min(n_p/2, N_o) + n_u \nonumber \\
& = \min(n_p/2, N_e) + \min(n_p/2, N_o) + \min(n_u, L).
\end{align}
\item{Finally let us consider the case $n_u < L^*$ and $n_u > L$.  Since $L <
n_u < L^* = L + t$, then  $L -t < n_u - t < L $ and $\min(n_u,
L)=L$. Therefore we can replace $n_u - t$ by $L$ in expression~(\ref{CasoReBajoTB})  if we substitute the sign $\le$ by $<$:}
\begin{align}
\label{CasoReBajoTBC2}  
\rank (\Y) & < \min(n_p/2, N_e) + \min(n_p/2, N_o) + L \nonumber \\
& = \min(n_p/2, N_e) + \min(n_p/2, N_o) + \min(n_u, L).
\end{align}
\end{enumerate}
\end{enumerate}

\noindent
Since we have covered all the possible cases, this shows that we can
write an upper limit on the rank of the masked case which does not depend on $R_e$
and $R_0$ as:
\begin{align}
\rank (\Y) & \le  \min(n_p/2, N_e)  + \min(n_p/2, N_o) + \min(n_u, L) . 
\label{FinalMascaraCasoB}
\end{align}
which proves the validity of the constraint~(\ref{RangoMascara}) given in section~\ref{sec:Mascara}.

\acknowledgments

We acknowledge partial financial support from the Spanish
\textit{Ministerio de Econom\'{\i}a y Competitividad}
Projects Consolider-Ingenio 2010
CSD2010-00064, ESP2015-70646-C2-1-R and AYA2015-64508-P (MINECO/FEDER,UE). This research was also supported by the RADIOFOREGROUNDS project, funded by the European Comission's H2020 Research Infrastructures under the Grant Agreement 687312. We also thank CSIC for a Proyecto Intramural Especial (201550I027). The \texttt{HEALPix} package~\cite{healpix} was used throughout the data analysis.







\end{document}